\documentclass{article}

\usepackage{graphicx}
\usepackage{enumerate}
\usepackage{natbib}
\usepackage{url} %

\usepackage{amsmath}
\usepackage{amssymb}
\usepackage{amsthm}
\usepackage{hyperref}
\usepackage{algorithm,algpseudocode}
\usepackage{dsfont}
\usepackage[usenames,dvipsnames]{xcolor} 
\usepackage{booktabs}
\usepackage[affil-it]{authblk}
\usepackage{cleveref}
\usepackage{tabularx}
\usepackage{enumitem}
\usepackage{stmaryrd}
\usepackage{bm}

\usepackage{acro}

\usepackage{pdflscape}

\newcommand{\blind}{0}

\theoremstyle{plain}
\newtheorem{theorem}{Theorem}

\usepackage{todonotes}
\usepackage{subfigure}
\usepackage{adjustbox}
\usepackage{multirow}
\usepackage{amsmath}
\usepackage{mathtools}
\usepackage{setspace}

\newcommand{\bs}{\boldsymbol}

\newcommand{\E}{\operatorname{E}}  
\newcommand{\N}{\mathcal{N}}  
\newcommand{\I}{\mathbf{I}} 
  
\newcommand{\T}{{\!\mathsf{T}}}  %

\newcommand{\x}{\mathbf{x}} 
 
\newcommand{\f}{\mathbf{f}} 
\newcommand{\y}{\mathbf{y}}  
\newcommand{\z}{\mathbf{z}}  
\newcommand{\K}{\mathbf{K}}

\newcommand{\C}{\mathbf{C}}
 
\renewcommand{\b}{\mathbf{b}}

\newcommand{\g}{\mathbf{g}} 
\renewcommand{\l}{\mathbf{l}} 
\renewcommand{\L}{\mathbf{L}}

\newcommand{\Q}{\mathbf{Q}}
\newcommand{\R}{\mathbf{R}}

\newcommand{\U}{\mathbf{U}}

\newcommand{\X}{\mathbf{X}} 
\newcommand{\Y}{\mathbf{Y}}
\newcommand{\Z}{\mathbf{Z}}

\newcommand{\indicator}[1]{\mathds{1}_{\{#1\}}}

\usepackage{comment}

\newif\ifshowcomments
\showcommentstrue   %

\ifshowcomments
  \includecomment{comment}
  
\else
\fi

\begin{document}

\def\spacingset#1{\renewcommand{\baselinestretch}%
{#1}\small\normalsize} \spacingset{1}

\title{Stochastic Variational Inference for Structured Additive Distributional Regression}

\if0\blind
{
  \author{Gianmarco Callegher (University of G\"ottingen), 
    Thomas Kneib (University of G\"ottingen), 
    Johannes S\"oding (Max-Planck-Institute for Multidisciplinary Sciences), 
    Paul F.V.\ Wiemann (The Ohio State University)}
} \fi

\maketitle

\bigskip

\begin{abstract}

Structured additive distributional regression extends generalized additive models by allowing all parameters of a response distribution to depend on structured additive predictors. Bayesian formulations regularize these models through prior distributions that enforce smoothness or shrinkage, but Markov chain Monte Carlo methods, the standard computational tool, scale poorly with sample size and require substantial runtime.
We develop a scalable stochastic variational inference framework for approximate Bayesian inference in structured additive distributional regression. Our key contribution is a variational family for the regression coefficients that is amortized over covariates, responses, and smoothing parameters. This construction allows the variational location and precision to adapt locally to the data, which leads to substantially higher ELBO values in markedly less computational time compared to existing approaches. To balance accuracy and scalability, we consider a block-structured variant that constrains the precision matrix but preserves essential dependence across smooth components. Both approaches are compared with a state-of-the-art dense variational approximation and with the widely used Integrated Nested Laplace Approximation (INLA).
We establish theoretical results on posterior propriety and local asymptotic Gaussianity that motivate the proposed Gaussian approximations, assess their performance in simulation studies involving logistic and gamma distributional regression, and demonstrate their scalability in an application to modeling injury counts in motor vehicle crashes in New York City with more than 200,000 observations. Across all settings, incorporating covariates, responses, and smoothing parameters into the construction of the variational distribution yields higher ELBO values in substantially less computational time compared with existing variational approaches.

\end{abstract}

\noindent%
{\it Keywords:} Bayesian Inference; Distributional Regression; Semiparametric Regression; Variational Bayes; Stochastic Variational Inference.
\vfill

\newpage
\spacingset{1.9} %

\section{Introduction}\label{sec:Introduction}

In recent years, interest has been increasing in flexible forms of regression models that overcome some of the limitations of classical forms of statistical modeling such as (generalized) linear models. Roughly, these developments can be grouped into (i) flexibility with respect to the chosen response distribution such that one can focus on other aspects than the conditional mean of the response variable \cite{Kneib2013Beyond,kneib2023rage}, (ii) flexibility with respect to the predictor specification going beyond linear combinations of covariate effects in semiparametric regression \cite{rupwan03,wood06}, (iii) allowing for dependence and unobserved heterogeneity between observations by including random or spatial effects \cite{BanCarGel03,fahkne11,fahrmeir2004penalized}. Many of these developments can be cast into the framework of structured additive distributional regression models \cite{kneib2023rage}, where the response distribution is specified based on a set of latent variables (e.g.\ the regression predictors or the regression coefficients) equipped with Gaussian (prior) distributions.

The complexity of such flexible forms of models demands well-suited inferential procedures. Bayesian inference based on \emph{Markov chain Monte Carlo (MCMC)} simulation techniques has proven very valuable in this regard (as can be seen from the abundance of Bayesian inferential procedures in the references above). This is owed on the one hand to the possibility of providing additional information to the model by including suitable forms of regularization priors that enforce desirable model properties such as shrinkage, smoothness or sparsity, and on the other hand to the implementation of divide-and-conquer strategies in MCMC, where large and complex models can be decomposed into smaller building blocks.

The main drawback of MCMC-based inference, however, is its high demand for computation time, especially when learning models with many parameters and/or large sample sizes. Other drawbacks include potentially slow mixing of the Markov chain, e.g.\ in case of highly correlated parameters, which further adds to the computation time, and the need to carefully monitor convergence in distribution over a high-dimensional parameter space. Therefore, fast and efficient \emph{approximate} Bayesian inference methods are growing in popularity. 

A very successful example is conducting inference using the \emph{Integrated Nested Laplace Approximation (INLA)}~\cite{rue2009approximate}. INLA has become extremely widely used in many statistical application areas due to the availability of fast, free software and the wide applicability of Latent Gaussian Models. In particular, INLA offers a wide range of modeling options for the (univariate) predictor of a regression model (relying on a structured additive predictor decomposition) and for considering dependence and unobserved heterogeneity with random and spatio-temporal effects. INLA gains much speed by adding the structured additive predictors to the set of Gaussian latent variables, which results in a sparse precision matrix of the joint prior for the latent variables. Consequently, this trick allows for the use of sparse matrix operations reducing its time complexity such that it can be applied to tasks with tens of thousands of latent variables.

\emph{Variational Inference (VI)} is a class of approximate Bayesian inference methods in which the posterior distribution is approximated by the best member of a pre-specified family of distributions \citep{Blei2017Variational,zhang2018advances}. Let $\z$ denote the latent variables and $\y$ the observed data. The posterior $p(\z \mid \y)$ is approximated by a \emph{variational distribution} $q(\z \mid \bs\phi)$, parameterized by $\bs\phi$. The variational parameters are obtained by maximizing the \emph{evidence lower bound (ELBO)} with respect to $\bs\phi$. Since the log marginal likelihood satisfies $\log p(\y) = \mathrm{ELBO}(\bs\phi) + \mathrm{KL}\!\left(q(\z \mid \bs\phi)\,\|\,p(\z \mid \y)\right)$, maximizing the ELBO is equivalent to minimizing the Kullback--Leibler divergence between $q(\z \mid \bs\phi)$ and the posterior $p(\z \mid \y)$, thereby approximating $p$ by $q$.

\emph{Stochastic Variational Inference (SVI)} extends VI to large-scale and complex models by optimizing the ELBO using stochastic gradient methods \citep{hoffman2013stochastic,zhang2018advances}. In SVI, gradients of the ELBO are estimated using Monte Carlo samples $\z^s \sim q(\z \mid \bs\phi)$ and, when applicable, mini-batches of observations. This yields scalable updates with per-iteration cost that does not grow linearly with the full dataset size. Moreover, because samples are drawn directly from $q(\z \mid \bs\phi)$ (rather than via a Markov chain), no mixing or burn-in is required, avoiding a common practical complication of MCMC.

Here, we develop stochastic variational inference for structured additive distributional regression models using multivariate Gaussian variational approximations for the latent regression coefficients. Building on recent Gaussian VI ideas for structured additive models \citep{kleinemeier2023scalable}, we propose a new construction tailored to distributional regression with smoothness penalties.

Our variational approximation factorizes into two multiplicative components: a likelihood term represented as a product of $N$ Gaussian factors (with $N$ the sample size) and an unnormalized prior term. This design is motivated by theoretical considerations on posterior propriety and asymptotic normality, and it enables efficient use of covariate information and smoothing parameters, which control the influence of the prior. We additionally introduce a second variational distribution for the smoothing parameters, parameterized by separate hyperparameters, and learn all variational parameters by stochastic maximization the ELBO.

The main methodological contributions of this work are twofold. First, we introduce an amortized variational family in which local Gaussian factors depend on the covariates, the response, and the smoothing parameters through a neural network mapping, leading to substantially faster ELBO improvement than static parametrizations. Second, we propose a block-diagonal parametrization that reduces the cost of matrix operations while retaining flexibility via a structured correction. Together with the variational treatment of the smoothing parameters, these components yield a scalable approach for joint inference in structured additive distributional regression.

We complement the methodological developments with theoretical results establishing sufficient conditions for posterior propriety and, under regularity conditions, consistency and asymptotic normality of the MAP together with a local Gaussian approximation of the posterior.

The structure of this article is as follows: \autoref{section:Structured additive distributional regression} introduces structured additive distributional regression and establishes theoretical results on posterior propriety and asymptotic Gaussianity. \autoref{section:Variational Distribution} presents our stochastic variational inference framework for structured additive distributional regression. \autoref{section:Simulation} presents a simulation study in which we compare our models with a dense Gaussian SVI baseline, MCMC and INLA. \autoref{section:Application} demonstrates the scalability of our approach in an application to modeling injury counts in motor vehicle crashes in New York City. In \autoref{section:Conclusion}, we summarize our results and briefly discuss limitations and potential directions for future research.

\section{Structured additive distributional regression (SADR)}\label{section:Structured additive distributional regression}

\subsection{Model Specification}\label{sec:model}

Given observational data $(y_i, \x_i)$, $i=1,\ldots,N$ of a response variable $y$ and covariates $\x$, distributional regression assumes that the conditional distribution of the responses given the covariates is modelled using a $P$-parametric probability density function (pdf) $p(y_i | \x_i, \bs\beta) = p(y_i | \bs\nu_i),$ where $\bs\nu_i = (\nu_{1}(\x_{i}, \bs\beta_1), \dots, \nu_{P}(\x_{i}, \bs\beta_P))^\T$ is the $P$-dimensional vector of parameters that characterizes the distribution (for example, in terms of location, scale and skewness). Each distributional parameter $\nu_p(\x_i, \bs\beta_p)$, $p=1,\ldots,P$, with $\bs\beta_p \in \mathbb{R}^{Q_p}$, is then related to a regression predictor $\eta_p(\x_i, \bs\beta_p)$ via a bijective response function $h_{p}(\cdot)$ as $\nu_{i, p} \equiv \nu_{p}(\x_i, \bs\beta_p) = h_p(\eta_{p}(\x_{i, p}, \bs\beta_p))$ to ensure appropriate restrictions on the parameter space of $\nu_p(\x_{i, p}, \bs\beta_p)$ (such as positivity for scale parameters). 

The regression predictor $\eta_{p}(\x_{i, p}, \bs\beta_p)$ is assumed to be of structured additive form, i.e.\ it is composed of $J_p$ functions as
\begin{align*}
    \eta_{i, p} \equiv \eta_{p}(\x_{i, p}, \bs\beta_p) = \sum_{j = 1}^{J_p} f_{p, j}(\x_{i, p, j}, \bs\beta_{p,j}),
\end{align*}
where $f_{p, j}(\x_i, \bs\beta_{p,j})$ are effects of different types such as nonlinear effects of continuous covariates, spatial effects, random effects, etc. and $\bs\beta_{p,j}$ are the corresponding regression coefficient subsets of $\bs\beta_p = (\bs\beta_{p,1}^\T,\ldots,\bs\beta_{p,J_p}^\T)^\T$. The different effect types can be represented through basis function expansion as $f_{p, j}(\x_i, \bs\beta_{p, j}) = \sum_{l=1}^{Q_{p, j}} \beta_{p, j, l}B_{p, j, l}(\x_i),$ where $B_{p, j, l}$ are known basis functions. 

As a major advantage, this setup offers considerable flexibility in choosing from a variety of response distributions and effect types which are all linear in the basis coefficients $\bs\beta_{p, j}$ such that each term $f_{p, j}$ of the structured additive predictor $\eta_p$ is associated with a design matrix $\X_{p, j}$ and the vector of function evaluations $\f_{p, j}=(f_{p, j}(\x_1, \bs\beta_{p, j}),\ldots,f_{p, j}(\x_N, \bs\beta_{p, j}))^\T$ can be written as $\f_{p, j} = \X_{p, j}\bs\beta_{p, j}$, where $\bs\beta_{p, j} = (\beta_{p, j, 1}, \dots, \beta_{p, j, Q_{p, j}})^\T \in \mathbb{R}^{Q_{p, j}}$.

To regularize estimation, we assume informative priors 
\begin{align*}
    p(\bs\beta_{p, j} | \lambda^2_{p, j}) \propto \exp\left\{ -\frac{\lambda^2_{p, j}}{2} \bs\beta_{p, j}^\T \K_{p,j} \bs\beta_{p, j} \right\},
\end{align*}
that implement prior assumptions such as smoothness or shrinkage with potentially rank-deficient precision matrix $\K_{p, j} \in \mathbb{R}^{Q_{p, j} \times Q_{p, j}}$ and precision parameter $\lambda^2_{p, j}$ governing the impact of the prior. The model specification is completed by assuming prior independence between different effects and  additional hyperpriors for $\lambda^2_{p, j}$, which depend on the chosen effect type and whether it is subject to effect selection or regularization. The joint prior for all basis coefficients $\bs\beta = (\bs\beta_1^\T,\ldots,\bs\beta_P^\T)^\T = (\bs\beta_{1,1}^T, \ldots, \bs\beta_{P,J_P}^T)^T \in \mathbb{R}^Q$ is then a normal distribution $\prod_{p=1}^P\prod_{j=1}^{J_p} \N\big(\bs\beta_{p,j} \big| \bs 0, \K_{p,j,\bs\lambda_{p,j}} \big) = \N\big(\bs\beta \big | \bs 0, \K_{\bs\lambda} \big)$ where the overall precision matrix $\K_{\bs\lambda}$ is block-diagonal with the matrices  $\lambda^2_{p, j} \K_{p, j}$ placed along the diagonal and $\bs\lambda$ is the vector of all precision parameters. 

\subsection{Posterior distribution}

Assuming conditional independence of the observations given the model parameters, the log-posterior distribution for all model parameters is given by
\begin{align}
    \ln p(\bs\beta, \bs\lambda | \y, \X) &=  %
    \sum_{i=1}^N \ln p( \y_i | \bs\nu_i) + \ln p(\bs\beta | \bs\lambda) + \ln p(\bs\lambda) + \text{const}
    \label{eq:p(beta, lambda | y, X)}
\end{align}
where $\bs\nu_i = (\nu_{1}(\x_{i}, \bs\beta_1), \dots, \nu_{P}(\x_{i}, \bs\beta_P))^\T$ as introduced above.
While often the posterior is accessed by MCMC, we will rely on stochastic variational inference (SVI) as described in detail in the next section. For constructing convenient and efficient approximation families for the regression coefficients, we will rely on multivariate normal distributions. In the following, we present two theoretical results supporting this choice.

The first result concerns the propriety of the posterior despite using partially improper normal distributions as priors for the different effects in SADR. For this, we are making the following assumptions:
\begin{enumerate}
\item[(T1-A)] Assumptions on the observation model: 
There exists an index set $\mathcal I^\ast\subseteq\{1,\ldots,N\}$ with $|\mathcal I^\ast|=n^\ast\geq1$ such that, for every $i\in\mathcal I^\ast$,
 \[
 \int\ldots\int  p(y_i | \bs\nu_i)d\eta_{i1}\ldots d\eta_{iP}<\infty\quad \text{(density integrable with respect to the predictors)}
 \]
 while $i\notin\mathcal I^\ast$, %
 \[
 p(y_i | \bs\nu_i)\le M\text{ for some finite }M \quad \text{(bounded density).}
 \]
\item[(T1-B)] Assumptions on the effect decomposition: The predictors can be reparameterised as
 \[
 \bs\eta_p = \U_p\bs\gamma_p + \Z_p\b_p + \Z_p^{\varepsilon}\b_p^{\varepsilon}
 \]
 where $\bs\gamma_p$ collects all parts of the effects with flat priors and design matrix $\U_p$, $\b_p^{\varepsilon}$ is the proper, normally distributed part of the effect with the largest basis dimension and design matrix $\Z_p^{\varepsilon}$ (without loss of generality, we assume that this is the last effect in the predictor), and $\b_p$ collects all remaining effects with proper, normal distributions and design matrix $\Z_p$ . 

Let $\U_{p,\mathcal I^\ast}$, $\Z_{p,\mathcal I^\ast}$, and $\Z^\varepsilon_{p,\mathcal I^\ast}$ denote the corresponding design matrices restricted to the observations in $\mathcal I^\ast$. We require that
(i) the design matrix $\U_{p,\mathcal I^\ast}$ has full rank $r_p$,
(ii) the composed matrix $(\U_{p,\mathcal I^\ast}, \Z_{p,\mathcal I^\ast})$ has rank $r_p + t_p$ with $t_p\le\dim\b_p$, and
(iii) the design matrix $\Z_{p,\mathcal I^\ast}^{\varepsilon}$ has full rank $k_{\varepsilon,p}$.
 We can then select a subset $\mathcal I_p\subseteq\mathcal I^\ast$ of $k_{\varepsilon,p}$ of observations such that the corresponding square submatrix $\Z^\varepsilon_{p,\mathcal I_p}$ is nonsingular, and normalize the predictors to
 \[
 \tilde{\bs\eta}_p = \tilde{\U}_p\bs\gamma_p + \tilde{\Z}_p\b_p + \b_p^{\varepsilon}.
 \]
 Note that the assumption on the integrability of the densities in (A) can be equivalently formulated in terms of the normalized predictors $\tilde{\bs\eta}_p$.
\item[(T1-C)] Assumptions on the prior precision matrices:
 \[
 \mathrm{rk}(\K_{p,j}) - \sum_{i=1}^{J_p-1}\mathrm{rk}(\K_{p,i}) + t_p -1 > 0, \quad j=1,\ldots,J_p-1 
 \]
\item[(T1-D)] Assumptions on predictor variation:
 \[
 \mathrm{SSE_p} = (\tilde{\bs\eta}_p - \tilde{\U}_p\bs\gamma_p - \tilde{\Z}_p\b_p)^T(\tilde{\bs\eta}_p - \tilde{\U}_p\bs\gamma_p - \tilde{\Z}_p\b_p)>0
 \]
\end{enumerate}

\begin{theorem}[Propriety of the posterior in SADR] In the model defined in Section~\ref{sec:model} and with either fixed precision parameters or proper priors for the precision parameters, assumptions (T1-A) to (T1-D) are sufficient to ensure that the posterior distribution is proper.\label{thm:propriety}
\end{theorem}

A sketch of the proof is presented in the appendix in Section~\ref{proof_theorem1}. Under stronger assumptions, we can also use improper priors for the hyperparameters $\bs\lambda$, see \citet[Section~3.3]{Klein2016Scale} for details. Note also that the result holds for finite samples, i.e.\ it does not require asymptotic considerations. In contrast, the following theorem shows that the maximum a posteriori (MAP) estimate in SADR is asymptotically normal and the posterior is locally Gaussian around the MAP. For this, let the posterior (up to normalization) be denoted as
$$
p_N(\bs\beta\mid Y, \X, \bs\lambda) \propto\ \exp \Big(\ell_N(\bs\beta) - \frac{1}{2} \bs\beta^\top \K_{\bs\lambda} \bs\beta\Big),
$$
where $\K_{\bs\lambda} \succeq 0$ and $\ell_N(\bs\beta) = \sum_{i=1}^N \log p(Y_i \mid \nu(\X_i, \bs\beta))$.

Define the MAP
$$
\hat{\bs\beta}_{\bs\lambda} = \arg\max_{\bs\beta \in \mathbb{R}^Q} \Big( \ell_N(\bs\beta) - \frac{1}{2} \bs\beta^\top \K_{\bs\lambda} \bs\beta \Big),
$$

We further define the observed information
$$
J_N(\bs\beta) = -\nabla^2\ell_N(\bs\beta),
$$
the observed posterior information
$$
H_N(\bs\beta) = -\nabla^2 \left[\ell_N(\bs\beta) - \frac{1}{2}\bs\beta^\top\K_{\bs\lambda}\bs\beta \right] = J_N(\bs\beta)+\K_{\bs\lambda},
$$
and the per-observation expected Fisher information
$$
I(\bs\beta) = -\mathbb E_{\X,Y} \left[\nabla^2 \log p(Y\mid\nu(\X,\bs\beta)) \right].
$$
All probability statements in the following are understood with respect
to the data-generating law $P_{\bs\beta_0}$. The assumptions used in
Theorem~\ref{thm:T2} are stated in Appendix~\ref{assumptions_theorem2}.

\begin{theorem}[MAP consistency and Laplace normality around the MAP]\label{thm:T2}
Under assumptions \ref{ass:t2-iid-spec} to \ref{ass:t2-integrated-tail-condition}, we obtain the following results:
\begin{enumerate}[label=\roman*.]
	\item \textbf{MAP consistency.} $\ \hat{\bs\beta}_{\bs\lambda} \xrightarrow{p} \bs\beta_0$.
	
    \item \textbf{Asymptotic Normality.}
      $\sqrt{N} \left(I(\hat{\bs\beta}_{\bs\lambda})\right)^{1/2}\left(\hat{\bs\beta}_{\bs\lambda} - \bs\beta_0\right) \xrightarrow{d} Z$
      where
      $Z \sim \mathcal{N}\big(\bm{0}, I_Q\big)$
      and therefore
      $\Vert \hat{\bs\beta}_{\bs\lambda} - \bs\beta_0 \Vert = O_p(N^{-1/2})$.

\item \textbf{Local quadratic expansion.}
For fixed $c>0$, let
$$
r_N=c\sqrt{\frac{\log N}{N}}
$$
and let
$$
B_N= \left\{ \bs\beta: \|\bs\beta-\hat{\bs\beta}_{\bs\lambda}\| \leq r_N \right\}.
$$
Then,
$$
\sup_{\bs\beta\in B_N}\left|\ell_N(\bs\beta) -\frac{1}{2}\bs\beta^\top \K_{\bs\lambda}\bs\beta - \left[\ell_N(\hat{\bs\beta}_{\bs\lambda}) -\frac{1}{2}\hat{\bs\beta}_{\bs\lambda}^\top \K_{\bs\lambda}\hat{\bs\beta}_{\bs\lambda} \right] + \frac{1}{2} (\bs\beta-\hat{\bs\beta}_{\bs\lambda})^\top H_N(\hat{\bs\beta}_{\bs\lambda}) (\bs\beta-\hat{\bs\beta}_{\bs\lambda}) \right| = o_p(1).
$$

For parts iv. and v., let $\Pi_N$ denote the posterior probability measure induced by $p_N(\bs\beta\mid\Y,\X,\bs\lambda)$. There exists a constant $c_0>0$ such that the following statements hold for every fixed $c>c_0$.

\item \textbf{Concentration.}

For fixed $c>c_0$, let
$$
B_N= \left\{\bs\beta: \Vert\bs\beta-\hat{\bs\beta}_{\bs\lambda}\Vert \leq c\sqrt{\frac{\log N}{N}} \right\}.
$$
Then,
$$
\Pi_N(B_N^c)\xrightarrow{p}0.
$$

\item \textbf{Local Gaussianity.}

For fixed $c>c_0$, let
$$
r_N=c\sqrt{\frac{\log N}{N}}
$$
and
$$
B_N= \left\{\bs\beta: \Vert\bs\beta-\hat{\bs\beta}_{\bs\lambda}\Vert \leq r_N \right\}.
$$
Then,
$$
\sup_{A\in\mathcal B(\mathbb R^Q)} \left| \Pi_N(A\cap B_N) - \Phi_N(A\cap B_N) \right| \xrightarrow{p}0,
$$
where $\mathcal B(\mathbb R^Q)$ denotes the Borel $\sigma$-algebra on $\mathbb R^Q$ and $\Phi_N$ is the Gaussian probability measure with mean $\hat{\bs\beta}_{\bs\lambda}$ and covariance $H_N(\hat{\bs\beta}_{\bs\lambda})^{-1}$.

\end{enumerate}
\end{theorem}

A proof is presented in the appendix in Section~\ref{proof_theorem2}. Together, Theorems~1 and 2 provide the theoretical foundation for the construction of the variational family in the next section.

\section{Stochastic variational inference for SADR}\label{section:Variational Distribution}

The posterior distribution of the regression coefficients and smoothing parameters in SADR is high dimensional and depends on several structured design matrices such that Markov chain Monte Carlo (MCMC) methods often become computationally expensive in this setting, especially for large sample sizes. We therefore develop stochastic variational inference (SVI) for SADR as a scalable, computationally efficient alternative. We construct multivariate Gaussian variational approximations for the regression coefficients, motivated by their asymptotic normality as presented in the last section. The smoothing parameters are incorporated into the variational approximation as well, allowing for joint inference and automatic adaptation to the degree of smoothness favoured by the data. The remainder of this chapter presents the construction of the variational distributions, explains how to estimate the smoothing parameters, describes the optimization of the variational parameters, outlines the mini-batching strategy used to ensure scalability, and concludes with an asymptotic complexity analysis.

\subsection{Amortized variational inference for the regression coefficients}\label{subsec:Amortized Variational Distribution}

Our variational distribution $q$ is composed of two factors. The first factor approximates the product of the $N$ $P$-dimensional likelihoods $p(y_i | \bs\eta_{i} = \X_{i} \bs\beta)$ over $\bs\beta$ by the product of $N$ $P$-dimensional Gaussian distributions $\N \big(\bs\eta_i|\bs\mu_i,(\L_i\L_i^\T )^{-1} \big)$. The second factor is simply the prior on $\bs\beta$. Consequently, the unnormalized variational distribution is given by:
\begin{align}
    \ln q(\bs\beta | \bs\lambda, \X, \bs\phi)
    &= -\frac{1}{2} \sum_{i=1}^N (\X_{i} \bs\beta - \bs\mu_i)^\T \L_i \L_i^\T (\X_{i} \bs\beta - \bs\mu_i) - \frac{1}{2} \bs\beta^\T \K_{\bs\lambda} \bs\beta + \text{const},
    \label{eq:q(beta|y,lambda,phi-improper)}
\end{align}
where $\X_{i} := \text{block-diag}(\x_{i,1}^\T, \ldots, \x_{i,P}^\T)$ are the covariates for data sample $i$ and the vectors $\bs\mu_i$ and the matrices $\L_i$, for $i = 1, \ldots, N$, are modelled using a neural network with weights and bias parameters $\bs\phi$ and input $y_i$, $\X_{i}$ and $\bs\lambda$, i.e.
\begin{align*}
    \bs\mu_i, \l_i = \text{NN}_{\bs\phi}(\X_i, y_i, \bs\lambda), \quad \text{for } i = 1, \ldots, N,
\end{align*}
where $\l_i = \psi(\L_i)$ is the log-Cholesky parametrization \citep{pinheiro1996unconstrained} of the matrix $\L_i$.
The normalized variational distribution of the latent variables can then be written as
\begin{equation}
    \ln q(\bs\beta | \bs\lambda, \X, \bs\phi)
    = -\frac{1}{2} \sum_{i=1}^N  \left( \bs\beta^\T \X_i^\T \L_i \L_i^\T \X_{i} \bs\beta - 2  \bs\beta^\T \X_{i}^\T \L_i \L_i^\T  \bs\mu_i \right) - \frac{1}{2} \bs\beta^\T \K_{\bs\lambda} \bs\beta + \text{const}
    \label{eq:q(beta|lambda,X,phi)}
    \end{equation}
i.e., as the log-density of the Gaussian distribution $\N \big( \bs\beta \big| \tilde{\bs\beta}_{\bs\lambda, \bs\phi} , \bs\Lambda_{\bs\lambda, \bs\phi}^{-1}\big)$
with
\begin{align}
    \tilde{\bs\beta}_{\bs\lambda, \bs\phi} = \bs\Lambda_{\bs\lambda, \bs\phi}^{-1} \sum_{i=1}^N \X_{i}^\T \L_i \L_i^\T \bs\mu_i \quad 
    \bs\Lambda_{\bs\lambda, \bs\phi} &= \sum_{i=1}^N  \X_i^\T \L_i \L_i^\T \X_{i} + \K_{\bs\lambda},
    \label{eq:beta_tilde,Lambda}
\end{align}
The precision matrix $\bs\Lambda_{\bs\lambda,\bs\phi}$ must be full rank for $q(\bs\beta \mid \bs\lambda,\X,\bs\phi)$ to be a proper Gaussian distribution. Rank deficiency can arise from non-identifiability of the additive predictor, for instance due to overlap between smooth components and intercept-like directions. To enforce identifiability and obtain full-rank design blocks, we apply the QR reparametrization described in \autoref{subsec:additive models}, which imposes the required constraints on the design matrices and thereby ensures that $\bs\Lambda_{\bs\lambda,\bs\phi}$ is non-singular.

\subsection{A block-diagonal variational distribution}\label{subsec:Amortized Variational Distribution - Block Diagonal}

The model proposed in \autoref{eq:beta_tilde,Lambda} necessitates computing the Cholesky decomposition of the matrix $\bs\Lambda_{\bs\lambda, \bs\phi}$ to solve the linear system needed to determine $\tilde{\bs\beta}_{\bs\lambda, \bs\phi}$ and to sample from $\N \big( \bs\beta \big| \tilde{\bs\beta}_{\bs\lambda, \bs\phi} , \bs\Lambda_{\bs\lambda, \bs\phi}^{-1}\big)$. This process has a complexity of $O(Q^3)$, rendering the model impractical for large numbers of regression coefficients. In this section, we introduce an alternative variational distribution which assumes independence between each smooth term $f_{i, p, j}$ initially and then reintroduces dependence using a new set of variational parameters. We employ a Gaussian distribution $\mathcal{N}\big(f_{i, p, j} = \x_{i, p, j}^\T\bs\beta_{p, j} | \mu_{i, p, j}, \sigma^2_{i, p, j} \big)$ to approximate each smooth term $f_{i, p, j}$, i.e.
\begin{align*}
    q(\bs\beta_{p, j} | \lambda_{p, j}, \X_{p, j}, \bs\phi_{p, j})
    = \mathcal{N} \big( \bs\beta_{p, j} \big | \tilde{\bs\beta}_{p, j}(\lambda_{p, j}, \bs\phi_{p, j}), \bs\Lambda^{-1}_{p, j}(\lambda_{p, j}, \bs\phi_{p, j})\big)
\end{align*}
with
\begin{align}
    \tilde{\bs\beta}_{p, j}(\lambda_{p, j}, \bs\phi_{p, j}) = \bs\Lambda^{-1}_{p, j}(\lambda_{p, j}, \bs\phi_{p, j}) \sum_{i=1}^N \frac{\mu_{i, p, j} \x_{i, p, j}}{\sigma^2_{i, p, j}}, \quad \bs\Lambda_{p, j}(\lambda_{p, j}, \bs\phi_{p, j}) &= \sum_{i=1}^N  \frac{\x_{i, p, j} \x_{i, p, j}^\T}{\sigma^2_{i, p, j}} + \lambda^2_{p, j}\K_{p, j},
    \label{eq:beta_pj_tilde,Lambda_pj}
\end{align}
where $\x_{i, p, j} \in \mathbb{R}^{Q_{p, j}}$ and $\mu_{i, p, j}$ and $\log\sigma_{i, p, j}$ are functions of $y_i$, $\x_{i, p, j}$ and $\lambda_{p, j}$:
\begin{align*}
    \mu_{i, p, j}, \log\sigma_{i, p, j} = \text{NN}_{\bs\phi_{p, j}}(\x_{i, p, j}, y_i, \lambda_{p, j}), \quad \text{for } i = 1, \ldots, N.
\end{align*}
The full variational distribution can be written as $q(\bs\beta | \bs\lambda, \X, \bs\phi) = \N \big( \bs\beta \big| \tilde{\bs\beta}_{\bs\lambda, \bs\phi} , \bs\Lambda_{\bs\lambda, \bs\phi}^{-1}\big)$, where $\tilde{\bs\beta}_{\bs\lambda, \bs\phi}$ is the vector obtained by stacking all $\tilde{\bs\beta}_{p, j}(\lambda_{p, j}, \bs\phi_{p, j})$ and $\bs\Lambda_{\bs\lambda, \bs\phi}^{-1}$ is a block-diagonal matrix obtained by stacking all $\bs\Lambda^{-1}_{p, j}(\lambda_{p, j}, \bs\phi_{p, j})$ along the diagonal.

To account for correlation beyond the block-diagonal structure, we introduce a correction structured lower triangular matrix $\L^{(c)}_{\bs\phi}$. The matrix $\L^{(c)}_{\bs\phi}$ contains non-zero sub-blocks strictly below the diagonal blocks of $\L_{\bs\phi, \bs\lambda}$, introducing off-diagonal dependencies between components. These sub-blocks are derived from the Cholesky factors of the local covariance matrices $\bs\Lambda^{-1}_{p, j}(\lambda_{p, j}, \bs\phi_{p, j})$. The resulting full variational distribution is:
\begin{align}
    q(\bs\beta | \bs\lambda, \X, \bs\phi) = \mathcal{N} \left(\tilde{\bs\beta}_{\bs\lambda, \bs\phi}, \left(\left(\L_{\bs\lambda, \bs\phi} + \L^{(c)}_{\bs\phi}\right)\left(\L_{\bs\lambda, \bs\phi} + \L^{(c)}_{\bs\phi}\right)^\T\right)^{-1}\right)
    \label{eq:q(beta|lambda,X,phi)_bd}
\end{align}
where $\L_{\bs\lambda, \bs\phi}$ is the Cholesky factor of the block-diagonal precision matrix $\bs\Lambda_{\bs\lambda, \bs\phi}$, i.e., $\bs\Lambda_{\bs\lambda, \bs\phi} = \L_{\bs\lambda, \bs\phi}\L_{\bs\lambda,\bs\phi}^\T$. The resulting structure allows for richer posterior dependencies while retaining computational tractability since the Cholesky decomposition complexity remains $O\big(Q^2 + \sum_p \sum_j Q_{p, j}^3\big)$.

\subsection{Estimation of the variational parameters}\label{sec:Estimation}

We optimize the variational parameters $\bs\phi$ by maximizing the evidence lower bound:
\begin{align}
    \text{ELBO}(\bs\phi) = \E_{q(\bs\beta | \bs\lambda, \X, \bs\phi)}\left[ \ln \frac{p(\y, \bs\beta | \bs\lambda)}{q(\bs\beta | \bs\lambda, \X, \bs\phi)}\right],
    \label{eq:elbo(phi)}
\end{align}
where $q(\bs\beta | \bs\phi, \bs\lambda, \X)$ can refer to either \autoref{eq:q(beta|lambda,X,phi)} or \autoref{eq:q(beta|lambda,X,phi)_bd}. To do that, we draw $S$ samples $\bs\beta^s \sim \N\big(\tilde{\bs\beta}_{\bs\lambda, \bs\phi}, \bs\Lambda_{\bs\lambda, \bs\phi}^{-1} \big)$ from $q(\bs\beta | \bs\lambda, \X, \bs\phi)$. This can be done by computing the Cholesky decomposition $\bs\Lambda_{\bs\lambda, \bs\phi} = \L_{\bs\phi, \bs\lambda}\L_{\bs\phi, \bs\lambda}^\T$, sampling $\bs\epsilon^s \sim \N(\bs 0, \I_{Q})$ and setting $\bs\beta^s = \g_{\bs\beta}(\bs\lambda, \X, \bs\epsilon^s, \bs\phi) := \tilde{\bs\beta}_{\bs\lambda, \bs\phi} + \L_{\bs\phi, \bs\lambda}^{-\T}\bs\epsilon^s$. We can now proceed by maximizing the ELBO with respect to $\bs\phi$ using the reparameterization trick \citep{kingma2013auto}, which consists of using the function $\g_{\bs\beta}(\bs\lambda, \X, \bs\epsilon, \bs\phi)$, which depends on $\bs\phi$, to replace the expectation over $\bs\beta$ by an expectation over $\bs\epsilon$:
\begin{align*}
    \E_{q(\bs\beta | \bs\lambda, \X, \bs\phi)} \left[ \ln \frac{p(\y, \bs\beta | \bs\lambda)}{q(\bs\beta | \bs\lambda, \X, \bs\phi)}  \right]
    &= \E_{\bs\epsilon} \left[ \ln \frac{p(\y, \g_{\bs\beta}(\bs\lambda, \X, \bs\epsilon, \bs\phi) | \bs\lambda)}{ q(\g_{\bs\beta}(\bs\lambda, \X, \bs\epsilon, \bs\phi) | \bs\lambda, \bs\phi) } \right]
\end{align*}
To optimize our parameters $\bs\phi$, we need to compute the gradient w.r.t. $\bs\phi$ of this expression:
\begin{align}
    \nabla_{\bs\phi} \E_{q(\bs\beta | \bs\lambda, \X, \bs\phi)} \left[ \ln \frac{p(\y, \bs\beta | \bs\lambda)}{q(\bs\beta | \bs\lambda, \X, \bs\phi)}  \right] &= \E_{\bs\epsilon} \left[ \nabla_{\bs\phi} \ln \frac{p(\y, \g_{\bs\beta}(\bs\lambda, \X, \bs\epsilon, \bs\phi) | \bs\lambda)}{ q(\g_{\bs\beta}(\bs\lambda, \X, \bs\epsilon, \bs\phi) | \bs\lambda, \bs\phi) } \right] \nonumber\\
    &\approx \frac{1}{S} \sum_{s = 1}^{S} \nabla_{\bs\phi} \left[ \ln p(\y, \bs\beta^s | \bs\lambda) - \ln q(\bs\beta^s | \bs\lambda, \X, \bs\phi) \right].
    \label{eq:grad_phi_beta E[(p/q)]}
\end{align}
A sketch of the training algorithm is described in \autoref{alg:train_MVGLM}. More sophisticated optimizers, such as Adam \citep{kingma2014adam}, may be used to update the variational parameters in line 3 for improved performance.
\begin{algorithm}[htb] 
\textbf{Input:} Training data: responses $(y_1, \ldots, y_N)$, covariates $(\X_1, \ldots, \X_P)$ and step-size $\alpha$
\begin{algorithmic}[1]
\State Initialize $\bs\phi$
\For{$t = 1, \ldots, T$}
    \State $\bs\phi^{t + 1} \gets \bs\phi^{t} + \alpha\nabla_{\bs\phi} \text{ELBO}(\bs\phi^t)$ using \autoref{eq:grad_phi_beta E[(p/q)]}
\EndFor
\State \textbf{return} $\tilde{\bs\beta}_{\bs\lambda,\bs\phi}$ and $\bs\Lambda_{\bs\lambda,\bs\phi}^{-1}$ %
\end{algorithmic}
\caption{Train SVI-DR by stochastic variational inference.}
\label{alg:train_MVGLM}
\end{algorithm}

\subsection{Estimation of the precision parameters}

In order to estimate the precision parameters $\bs\lambda$, we  include them into the set of latent variables and sample them employing the variational approximation in \autoref{eq:q(beta|y,lambda,phi-improper)}, including the precision parameters, to approximate \autoref{eq:p(beta, lambda | y, X)}. This results in 
\begin{align*}
    q(\bs\beta, \bs\lambda | \X, \bs\phi)
    &= q(\bs\beta | \bs\lambda, \X, \bs\phi_{\bs\beta}) \cdot q(\bs\lambda | \bs\phi_{\bs\lambda}).
\end{align*}
Assuming independence among all smoothing parameters $\lambda_{p, j}$, we obtain
\begin{align*}
    q(\bs\lambda | \bs\phi_{\bs\lambda}) = \prod_{p = 1}^{P}\prod_{j = 1}^{J_p} q(\lambda_{p, j} | \bs\phi_{\lambda_{p, j}})
\end{align*}
with, for example, log-normal distributions for each $q(\lambda_{p, j} | \bs\phi_{\lambda_{p, j}})$. In that case, the vector of variational parameters $\bs\phi_{\bs\lambda}$ contains all the log-location and log-scale parameters that characterize each variational log-normal distribution $q(\lambda_{p, j} | \bs\phi_{p, j})$.
 
We then optimize the variational parameters $\bs\phi=(\bs\phi_{\bs\beta}, \bs\phi_{\bs\lambda})$ by maximizing
\begin{align}
    \text{ELBO}(\bs\phi) = \E_{q(\bs\beta, \bs\lambda | \X, \bs\phi)} \left[ \ln p(\y, \bs\beta, \bs\lambda)  \right] - \E_{q(\bs\beta, \bs\lambda | \X, \bs\phi)} \left[ \ln q(\bs\beta, \bs\lambda | \X, \bs\phi) \right]
    \label{eq:elbo(phi,lambda)}
\end{align}
again using the reparametrization trick. First, we draw $S_{\lambda}$ samples $\bs\lambda^{s_{\lambda}}$ from $q(\bs\lambda | \bs\phi_{\bs\lambda})$. Each $\bs\lambda^{s_{\lambda}}$ maps to a different variational distribution $\N\big(\tilde{\bs\beta}_{\bs\lambda^{s_{\lambda}}, \bs\phi}, \bs\Lambda_{\bs\lambda^{s_{\lambda}}, \bs\phi}^{-1} \big)$ from which we draw $S_{\beta}$ samples of $\bs\beta^s = \g_{\bs\beta}(\bs\lambda^{s_{\lambda}}, \X, \bs\epsilon, \bs\phi_{\bs\beta})$. To optimize our parameters $\bs\phi$, we need to compute the gradient w.r.t. $\bs\phi$ of \autoref{eq:elbo(phi,lambda)} as
\begin{align}
    \nabla_{\bs\phi} \E_{q(\bs\beta, \bs\lambda | \X, \bs\phi)} \left[ \ln \frac{p(\y, \bs\beta, \bs\lambda)}{q(\bs\beta, \bs\lambda | \X, \bs\phi)}  \right] \approx \frac{1}{S_{\lambda} \cdot S_{\beta}} \sum_{s_{\bs\lambda} = 1}^{S_{\lambda}} \sum_{s_{\bs\beta} = 1}^{S_{\beta}} \nabla_{\bs\phi} \left[ \ln p(\y, \bs\beta^{s_{\beta}}, \bs\lambda^{s_{\lambda}}) - \ln q(\bs\beta^{s_{\beta}}, \bs\lambda^{s_{\lambda}} | \X, \bs\phi) \right].
    \label{eq:grad_phi_beta,lambda E[(p/q)]}
\end{align}

The estimation algorithm is similar to \autoref{alg:train_MVGLM}, where, instead of updating the variational parameters using \autoref{eq:grad_phi_beta E[(p/q)]} (line 3), one has to use \autoref{eq:grad_phi_beta,lambda E[(p/q)]}.

\subsection{Mini-batching}

Mini-batching enables scalable variational inference by reducing the computational cost per iteration. In our amortized variational distribution, the local parameters $\bs\mu_i$ and $\L_i$ depend on the covariates, so the variational distribution must itself be constructed from the mini-batch. This requires unbiased mini-batch estimators of the sufficient statistics defining the Gaussian variational distribution.

Let $\mathcal{B}$ be a mini-batch of size $B$ sampled at random from the full dataset of size $N$. We construct unbiased estimators of the precision matrix and the natural mean as
\begin{align*}
\hat{\bs\Lambda}_{\bs\lambda, \bs\phi}^{(\mathcal{B})}
= \frac{N}{B} \sum_{i \in \mathcal{B}} \X_i^\top \L_i \L_i^\top \X_i + \K_{\bs\lambda}, \quad \hat{\bs\Lambda}_{\bs\lambda, \bs\phi}^{(\mathcal{B})}\,\hat{\tilde{\bs\beta}}_{\bs\lambda, \bs\phi}^{(\mathcal{B})}
= \frac{N}{B} \sum_{i \in \mathcal{B}} \X_i^\top \L_i \L_i^\top \bs\mu_i.
\end{align*}

These scaled sufficient statistics are unbiased estimators of their full-data counterparts, \textit{i.e.}
\begin{align*}
\mathbb{E}_{\mathcal{B}}\!\left[
\hat{\bs\Lambda}_{\bs\lambda, \bs\phi}^{(\mathcal{B})}
\right]
= \bs\Lambda_{\bs\lambda, \bs\phi},
\qquad
\mathbb{E}_{\mathcal{B}}\!\left[
\hat{\bs\Lambda}_{\bs\lambda, \bs\phi}^{(\mathcal{B})}\,
\hat{\tilde{\bs\beta}}_{\bs\lambda, \bs\phi}^{(\mathcal{B})}
\right]
= \bs\Lambda_{\bs\lambda, \bs\phi}\,\tilde{\bs\beta}_{\bs\lambda, \bs\phi},
\end{align*}
However, because the variational distribution depends nonlinearly on the sampled mini-batch, the resulting mini-batch ELBO and, in general, its gradient are not unbiased estimators of their full-data counterparts. A detailed derivation and discussion are provided in Appendix~\ref{app:mini-batching}.

\subsection{Asymptotic complexity}

\autoref{tab:asymptotic-complexity} in the Appendix compares the per-iteration asymptotic complexity of several approaches for estimating the regression coefficients $\bs\beta$. In addition to our amortized variational construction in \autoref{subsec:Amortized Variational Distribution} and its block-diagonal variant in \autoref{subsec:Amortized Variational Distribution - Block Diagonal}, we include a dense Gaussian SVI baseline based on \citet{kleinemeier2023scalable} (learning a full precision matrix rather than using a factor structure), as well as INLA and the No-U-Turn Sampler \citep[NUTS;][]{hoffman2014no}. These competitors are considered throughout the simulation study in \autoref{section:Simulation}, which motivates their inclusion in the asymptotic analysis.

Due to the QR reparametrization applied to each design matrix $\X_i$, the precision matrix in INLA becomes dense, leading to cubic complexity for its factorization. We employ one NUTS run for each coefficient block $\bs\beta_p$. Assuming $N<Q$, the dominant cost for the variational models in \autoref{subsec:Amortized Variational Distribution} and \autoref{subsec:Amortized Variational Distribution - Block Diagonal} is the Cholesky factorization of $\bs\Lambda_{\bs\lambda,\bs\phi}$, and the block-diagonal construction substantially reduces this burden. The dense Gaussian baseline of \citet{kleinemeier2023scalable} attains the optimal asymptotic complexity $\mathcal{O}(Q^2)$. While comparing INLA with SVI purely in terms of asymptotic complexity is difficult, INLA can be accelerated considerably by exploiting sparsity, reducing factorization to $\mathcal{O}(Q^{3/2})$ \citep{rue2017bayesian}. Conversely, the main computational bottleneck of NUTS is the (data- and geometry-dependent) number of leapfrog steps, which can increase substantially.

Although INLA and the dense Gaussian baseline are asymptotically more favorable per iteration, the empirical results in \autoref{section:Simulation} show that the proposed amortized variants reach high ELBO values in fewer optimization epochs and with less wall-clock time, yielding faster convergence in practice.

\section{Simulations}\label{section:Simulation}

\subsection{Goals of the simulation study}

We aim to evaluate the ability of our SVI approach to approximate the posterior in structured additive distributional regression accurately. We consider two different situations: (i) Binary responses whose likelihood terms are notoriously hard to approximate by normal distributions~\cite{albert1993bayesian} and (ii) gamma responses for which both the location and the scale parameter depend on covariates as an example of distributional regression with skewed response distribution. While the skewness makes the likelihood terms hard to fit using our variational normal distributions, the complex dependence between location and scale of the gamma distribution also challenges the SVI variant with block-diagonal covariance structure introduced in \autoref{subsec:Amortized Variational Distribution - Block Diagonal}.

In both cases, we consider MCMC with a large number of posterior samples as the gold standard against which we benchmark the approximation to the posterior provided by SVI. To quantify the deviation between the approximate posterior and the posterior derived from MCMC, we rely on the Wasserstein Distance (WD), also known as the Earth Mover's Distance (EMD) or the Mallows distance \citep{villani2008optimal}. The Wasserstein distance is a metric that can be used to quantify the difference between two probability distributions in a metric space. Let $p$ and $q$ be two probability distributions defined on a metric space $(M, d)$, where $M$ is a set and $d$ is the metric. Let $\Omega(p, q)$ denote the set of all joint distributions on $M \times M$ with marginals $p$ and $q$. The $r$-th Wasserstein distance between $p$ and $q$ for $r \in [1, +\infty)$ is then defined as
\begin{align*}
    W_r(p, q) = \left(\text{inf}_{\omega \in \Omega(p, q)}\E_{(\theta_p, \theta_q) \sim \omega} \big[ d(\theta_p, \theta_q)^r \big] \right)^{1 / r}.
\end{align*}

In our analyses, we use the 2-Wasserstein distance, $W_2$, with the Euclidean distance as the underlying metric. For its computation, we use the Python Optimal Transport (POT) library \citep{flamary2021pot}.

\subsection{Model specification}

We compare our two proposed approaches with a dense Gaussian SVI baseline and Integrated Nested Laplace Approximations (INLA), using MCMC as the reference posterior. INLA \citep{rue2017bayesian} is a widely used framework for approximate Bayesian inference that exploits nested Laplace approximations to efficiently approximate posterior distributions. Its broad adoption is supported by efficient and freely available software and its applicability to a wide range of latent Gaussian models \citep{lindgren2015bayesian}. The dense Gaussian SVI baseline is inspired by \citet{kleinemeier2023scalable} and provides a variational benchmark without the proposed amortized construction. In all approaches, in accordance with the specifications in INLA, we parameterize the smoothing parameters on the log scale,
\begin{align*}
    \bs\tau = (\log\lambda_{1, 1}, \ldots, \log\lambda_{1, J_p}, \ldots, \log\lambda_{P, 1}, \ldots, \log\lambda_{P, J_P})^\T
\end{align*}
where $\lambda_{p,j}^2$ is the corresponding precision.

This leads us to the following five competing approaches:
\begin{enumerate}
    \item \texttt{svi\_amortized} --  SVI with amortized variational distribution as presented in \autoref{subsec:Amortized Variational Distribution} where the whole precision matrix is estimated exploiting the covariates $\X$, the smoothing parameters $\bs\tau$ and the observations $\y$.
    \item \texttt{svi\_amortized\_bd} -- SVI with amortized variational distribution and block-diagonal dependence structure as presented in \autoref{subsec:Amortized Variational Distribution - Block Diagonal}.
    \item \texttt{svi\_dense} -- SVI where the variational distributions for the regression coefficients $\bs\beta$ and the smoothing parameters $\bs\tau$ are assumed to be independent. This model is inspired by~\cite{kleinemeier2023scalable} but, instead of employing a factor structure for the precision matrix for the coefficients $\bs\beta$, we learn a full matrix. The variational distribution for $\bs\beta$ is a full multivariate normal distribution parametrized by the location vector $\tilde{\bs\beta}$ and the Cholesky factor $\L$:
    \begin{align*}
    q(\bs\beta) = \mathcal{N} \left(\tilde{\bs\beta}, \left(\L\L^\T\right)^{-1}\right).
    \end{align*}
    The variational parameters $\bs\phi$ are simply $\tilde{\bs\beta}$ and $\l$, where $\l = \psi(\L)$ is the log-Cholesky parametrization of the matrix $\L$.
    \item \texttt{inla} - Integrated Nested Laplace Approximation \citep{rue2017bayesian}. 
    \item \texttt{mcmc} - Markov Chain Monte Carlo. We used the software Liesel~\citep{liesel} with one NUTS sampler \citep{hoffman2014no} for each block of the regression coefficients $\bs\beta$ and another one for all smoothing parameters $\bs\tau$.
\end{enumerate}

A detailed description of the neural architectures, priors, Monte Carlo settings, and optimizer specifications used in all SVI models is provided in Appendix~\ref{app:architecture-sim}. To monitor convergence of the variational optimization, we track a smoothed estimate of the ELBO based on an exponential moving average of the mini-batch ELBO values. This MA--ELBO provides a low-variance diagnostic suitable for stochastic optimization. A detailed description is given in Appendix~\ref{subsec:ma_elbo}.

\subsection{Binary regression}

In this section, we consider logistic regression models with binary responses and correlated sparse-region covariates, which can lead to biased estimates and high uncertainty~\citep{HasTibFri09}, providing a challenging case for the normal distributions that we utilize in the construction of the variational family. For each of the 100 simulation replications, we generate $N=200$ observations according to the following scheme:
\begin{itemize}
    \item Generate $x_{i, 1}$ from a three-component mixture of $\mathcal{U}(0, \frac{\pi}{3})$, $\mathcal{U}(\frac{\pi}{3}, \frac{2\pi}{3})$, and $\mathcal{U}(\frac{2\pi}{3}, \pi)$ with mixture weights $9/20$, $2/20$, and $9/20$.
    \item Generate $z_{i, 2}$ from a two-component mixture of $\mathcal{U}(-\pi, -\frac{\pi}{3})$ and $\mathcal{U}(-\frac{\pi}{3}, 0)$ with mixture weights $18/20$ and $2/20$.
    \item Generate $x_{i, 2} = (\rho * x_{i, 1}) + \sqrt{1 - \rho^2} * z_{i, 2}$ where $\rho = -0.7$ governs the correlation between $x_1$ and $x_2$.
    \item Generate $y_i \sim \text{Bernoulli}(\text{logit}^{-1}(\sin(1.75 * x_{i, 1}) + \cos(-1.75 * x_{i, 2})))$.
\end{itemize}

For each simulation replication, we change the data and initialize all SVI models and MCMC with a different seed.

\begin{figure}[t]
    \centering
    \includegraphics[width=0.45\textwidth]{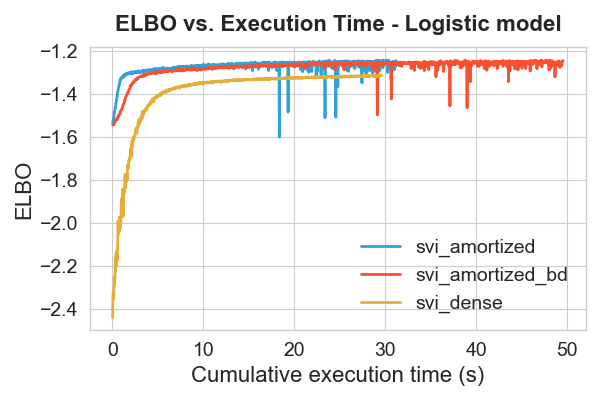}
    \includegraphics[width=0.45\textwidth]{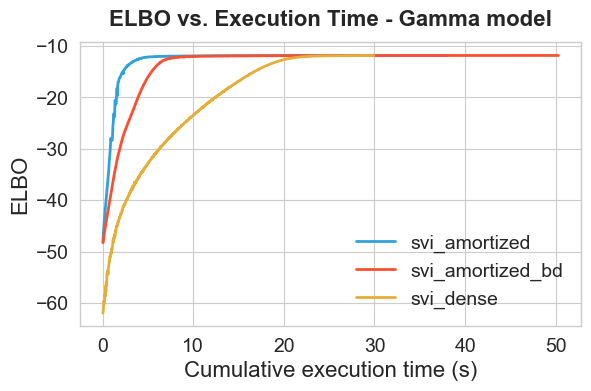}
    \caption{Relationship between execution time (in seconds) and ELBO after 1000 epochs for each SVI model in the logistic model (left) and the Gamma model (right). In both simulation settings, \texttt{svi\_dense} converges more slowly and reaches lower ELBO values, whereas \texttt{svi\_amortized} and \texttt{svi\_amortized\_bd} attain higher ELBOs in substantially less CPU time, emphasizing the efficiency of amortized inference.}
    \label{fig:execution_time_vs_elbo}
\end{figure}

\autoref{fig:execution_time_vs_elbo} illustrates the execution time (in seconds) as a function of the ELBO after 1000 epochs for each SVI model in the logistic and Gamma simulation settings. The \texttt{svi\_dense} model converges more slowly and attains a lower ELBO, whereas both \texttt{svi\_amortized} and \texttt{svi\_amortized\_bd} achieve higher ELBO values in less CPU time. These results highlight the computational efficiency of amortized inference compared to the dense full-precision approach.

\begin{figure}[t]
    \centering
    \includegraphics[width=0.75\textwidth]{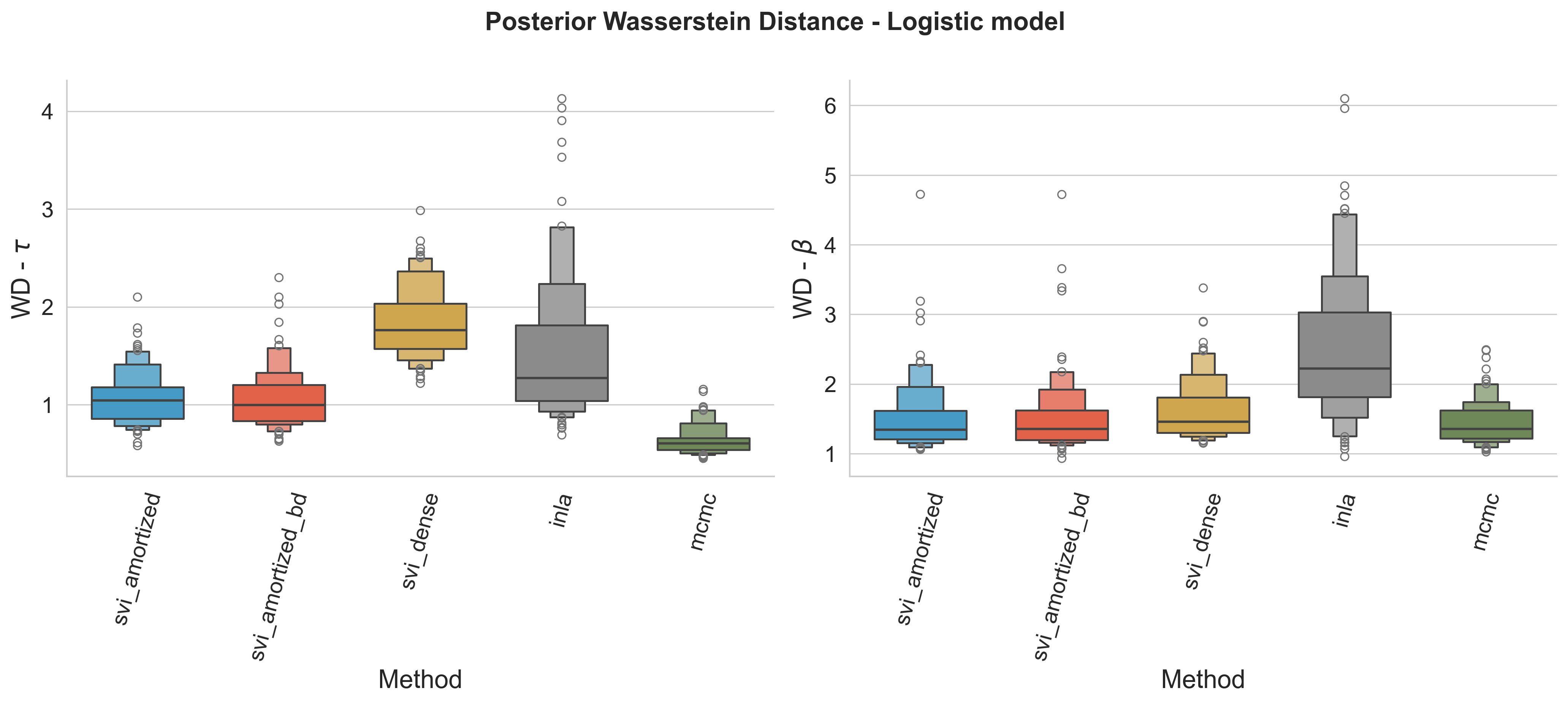}

    \vspace{0.4cm}

    \includegraphics[width=0.75\textwidth]{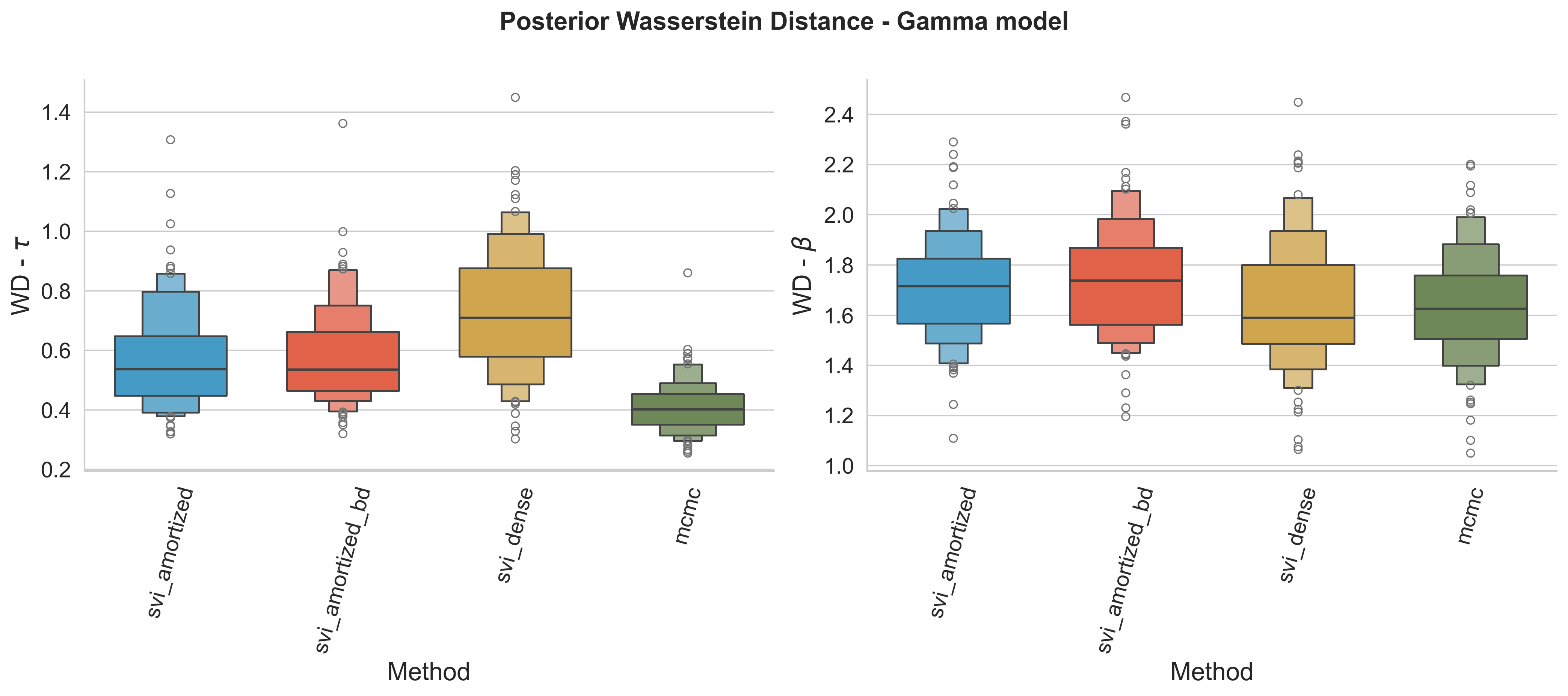}
    \caption{WD between the posterior distributions determined from MCMC and the approximate posterior obtained by each of the SVI approaches, INLA when available, and an independent MCMC posterior estimate in the logistic model (top) and the Gamma model (bottom). Each row reports the WD for $\bs\beta$ and $\bs\tau$. Each simulation replicate produces one sample of WD from which the boxen-plots are calculated. INLA is missing for the Gamma model since it cannot estimate both parameters of the Gamma.}
    \label{fig:wd-posterior}
\end{figure}

The top row of \autoref{fig:wd-posterior} assesses the ability of the variational families to approximate the posterior at the end of the estimation process, comparing them with INLA, where available, and a reference MCMC chain. The MCMC results provide a graphical benchmark for convergence, since the WD lacks an absolute scale. In the logistic regression setting, all SVI models produce results comparable to MCMC for the regression coefficients, whereas INLA shows poorer performance, with both a higher central tendency and greater dispersion. For the precision parameters $\bs\tau$, our models remain consistent with MCMC, while \texttt{svi\_dense}, and even more so INLA, exhibit increased location and variability.

In addition to evaluating posterior approximation quality via the WD, we also assessed the coverage probabilities of the estimated credible intervals for the smooth effects in the Bernoulli regression model. \autoref{fig:bernoulli-coverage-probabilities}, in the Appendix, presents the pointwise coverage of the nominal 95\% credible intervals across the domain of the two covariates $x_{1}$ (left panel) and $x_{2}$ (right panel). Overall, all SVI variants achieve coverage patterns comparable to those obtained by MCMC, which we treat as the gold standard. The amortized models (\texttt{svi\_amortized} and \texttt{svi\_amortized\_bd}) align most closely with MCMC across the entire covariate domain, demonstrating stable performance even in regions where the underlying nonlinear effect exhibits steep changes. By contrast, \texttt{svi\_dense} and INLA show slightly lower coverage in some intervals, particularly in the vicinity of sharp transitions. Despite these local deviations, the general agreement among all methods indicates that our proposed amortized approaches provide reliable uncertainty quantification in logistic distributional regression.

\subsection{Gamma regression}

In this second simulation scenario, we consider a distributional regression setup with gamma-distributed responses obtained from the following scenario:
\begin{itemize}
    \item Generate $x_i \sim \mathcal{U}(0, \pi)$.
    \item Compute $\mu_i = \exp{\{1.3 + 0.7 \sin(-x_i)\}}$ and $\sigma_i^2 = \exp{\{ -0.4 -0.9\cos(x_i)\}}^2$.
    \item Generate $y_i \sim \text{Gamma}(\mu_i, \sigma_i^2)$ where $\mu_i$ is the mean of the Gamma distribution and $\sigma_i^2$ is the variance.
\end{itemize}

For each of the 100 simulation replications, we generate $N=1000$ observations and initialize all SVI models and MCMC with a different seed. Note that in this scenario, we could not use the software INLA, since there are no APIs for estimating both moments of a Gamma response variable.

For the Gamma regression model, the execution-time results in \autoref{fig:execution_time_vs_elbo} are consistent with those observed for the logistic model. In this case, the gap between \texttt{svi\_dense} and the amortized models (\texttt{svi\_amortized} and \texttt{svi\_amortized\_bd}) is more pronounced, highlighting how the amortized approaches achieve higher ELBO values in significantly less time.

For the Gamma regression model, the bottom row of \autoref{fig:wd-posterior} shows the Wasserstein distance between the posterior distributions estimated by each model and the MCMC posterior at the end of the estimation process. As in the logistic regression setting, all SVI variants achieve comparable performance in terms of both median and variability. For the smoothing parameters $\bs\tau$, our approach more closely reproduces the MCMC posterior than \texttt{svi\_dense}.

\autoref{fig:gamma-coverage-probabilities} presents the pointwise coverage across the domain of the covariate $x$, with the left panel corresponding to the mean parameter $\mu$ 
and the right panel to the variance parameter $\sigma^{2}$. 

Across both response parameters, the SVI methods reproduce the general coverage  patterns obtained from MCMC, although differences between model variants are more  pronounced than in the logistic regression case. In particular, \texttt{svi\_amortized} and \texttt{svi\_amortized\_bd} again provide the closest alignment with the MCMC benchmark, whereas \texttt{svi\_dense} exhibits more systematic deviations, especially in regions where $\mu$ and $\sigma^{2}$ vary simultaneously. This behaviour reflects the increased difficulty of capturing dependence between the two distributional parameters in Gamma regression. Overall, the results indicate that the amortized approaches are better suited for reliable uncertainty quantification in distributional regression models with skewed responses.

\section{Applications}\label{section:Application}

We apply the proposed stochastic variational inference approach to model the number of injured persons in motor vehicle crashes recorded in New York City. The data come from the open dataset published by the New York Police Department, covering all reported collisions from 2023 to 2025 and comprising approximately 219{,}000 observations after cleaning and filtering incomplete records. Preliminary analyses with a Negative Binomial response indicated only mild overdispersion and no systematic covariate effects on the dispersion parameter, so in this section we present results for a Poisson model.

Let $y$ denote the number of injured persons in a crash, assumed to follow a Poisson distribution with mean $\mu$. The mean is linked to a structured additive predictor through a log-link,
\begin{align*}
  \log(\mu)
  &= f_{\mu,1}(\text{min}) 
   + f_{\mu,2}(\text{dow}) 
   + f_{\mu,3}(\text{month}) 
   + \beta_{\text{year}}\,\text{year} 
   + f_{\mu,4}(\text{lat}) 
   + f_{\mu,5}(\text{lon}) 
   + f_{\mu,6}(\text{lat},\text{lon}).
\end{align*}

The covariates include the crash time in minutes after midnight (min), day of week (dow), month of year (month), calendar year (year), geographic coordinates (lat, lon). All smooth components are specified through penalized B-splines. Marginal spatial variation in latitude and longitude is modeled with univariate B-splines, each comprising 20 coefficients, while their joint spatial effect, $f_{\mu,6}(\text{lat},\text{lon})$, is captured by an anisotropic tensor-interaction spline with $6\times 6$ coefficients. The variable \emph{min} (minutes after midnight) is represented by a cyclic spline to guarantee smoothness across the transition from 23{:}59 to 00{:}00, and cyclic splines are likewise used for day of the week and month. The effect of \emph{year} enters the model linearly.

In contrast to the simulation study, where we also compared against INLA and MCMC reference implementations, the real-data analysis focuses solely on the three SVI schemes due to the size of the data set: our fully amortized model (\texttt{svi\_amortized}), its block-diagonal precision variant (\texttt{svi\_amortized\_bd}), and a non-amortized dense Gaussian variational approximation (\texttt{svi\_dense}). A detailed description of the neural architectures, hyperpriors, and optimization settings used for these SVI models is provided in Appendix~\ref{app:architecture}. In~\ref{app:second_application} we worked with a smaller data set to make MCMC benchmarking feasible and assessed the accuracy of our methods not only for the mean but also for dispersion and other distributional components.

As in the simulation study, convergence of the variational inference procedure is assessed using the MA--ELBO, an exponential moving average of the noisy mini-batch ELBO estimates. This stabilised diagnostic allows us to determine when the optimization has effectively converged. See Appendix~\ref{subsec:ma_elbo} for details.

Figure~\ref{fig:elbo_nyc} in the Appendix displays the estimated ELBO trajectories over optimization time after 100 epochs. Both amortized variants quickly attain substantially higher ELBO values than the dense variational baseline, and they do so in markedly less time. The \texttt{svi\_amortized\_bd} variant exhibits a brief dip in the early iterations, but it soon recovers and closely tracks the fully amortized model, while \texttt{svi\_dense} converges more slowly.

\begin{figure}[htbp]
  \centering
  \includegraphics[width=\textwidth]{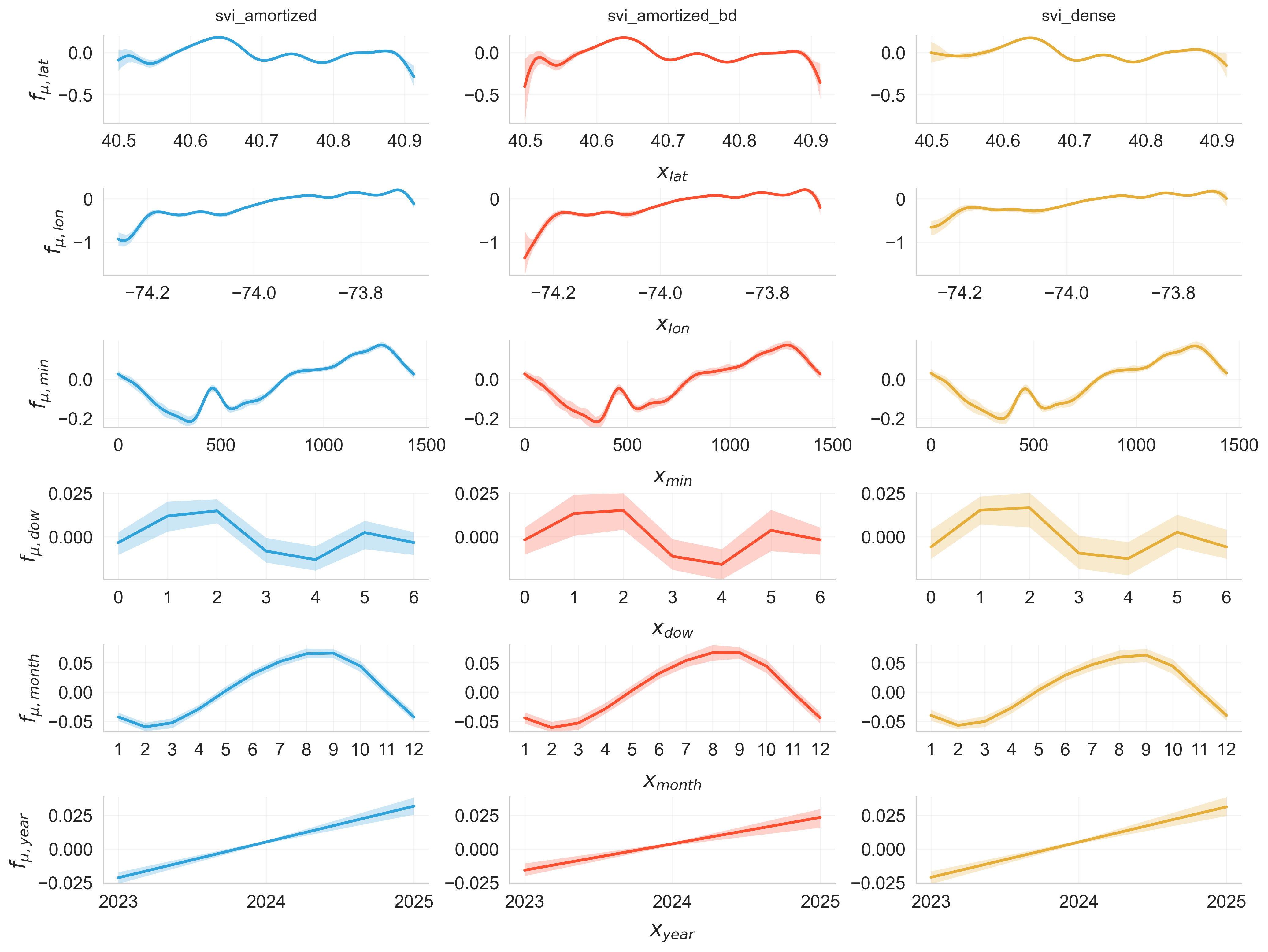}
  \caption{Posterior means (solid lines) and pointwise credible bands (shaded areas) of selected smooth effects for the three SVI methods: \texttt{svi\_amortized} (left column), \texttt{svi\_amortized\_bd} (middle column), and \texttt{svi\_dense} (right column). Rows correspond to latitude, longitude, minutes after midnight, day of the week, month and year respectively. Overall, the estimated effects are very similar across methods. The largest discrepancies occur for the spatial effects at the left-hand side of the domain in latitude and longitude, where the dense baseline deviates slightly from the two amortized variants, while temporal effects in \emph{min}, \emph{dow}, and \emph{month} are nearly indistinguishable.}
  \label{fig:effects_nyc}
\end{figure}

Figure~\ref{fig:effects_nyc} compares the estimated smooth components for latitude, longitude, minutes after midnight, day of the week, month and year. Across all three methods, the shapes of the effects are broadly similar, with overlapping credible bands over most of the covariate ranges. Differences are most pronounced in the spatial effects at the left-hand side of the latitude–longitude domain, where \texttt{svi\_dense} shows slightly different behaviour compared with the amortized SVI models. In contrast, the temporal effects in \emph{min}, \emph{dow}, and \emph{month} are virtually identical, indicating that all SVI schemes recover the same main structure of the mean response. Combined with the ELBO and timing results in Figure~\ref{fig:elbo_nyc}, this suggests that the proposed amortized SVI provides a more computationally efficient approximation than the dense baseline while preserving the qualitative interpretation of the effects.

\begin{figure}[htbp]
  \centering
  \includegraphics[width=\textwidth]{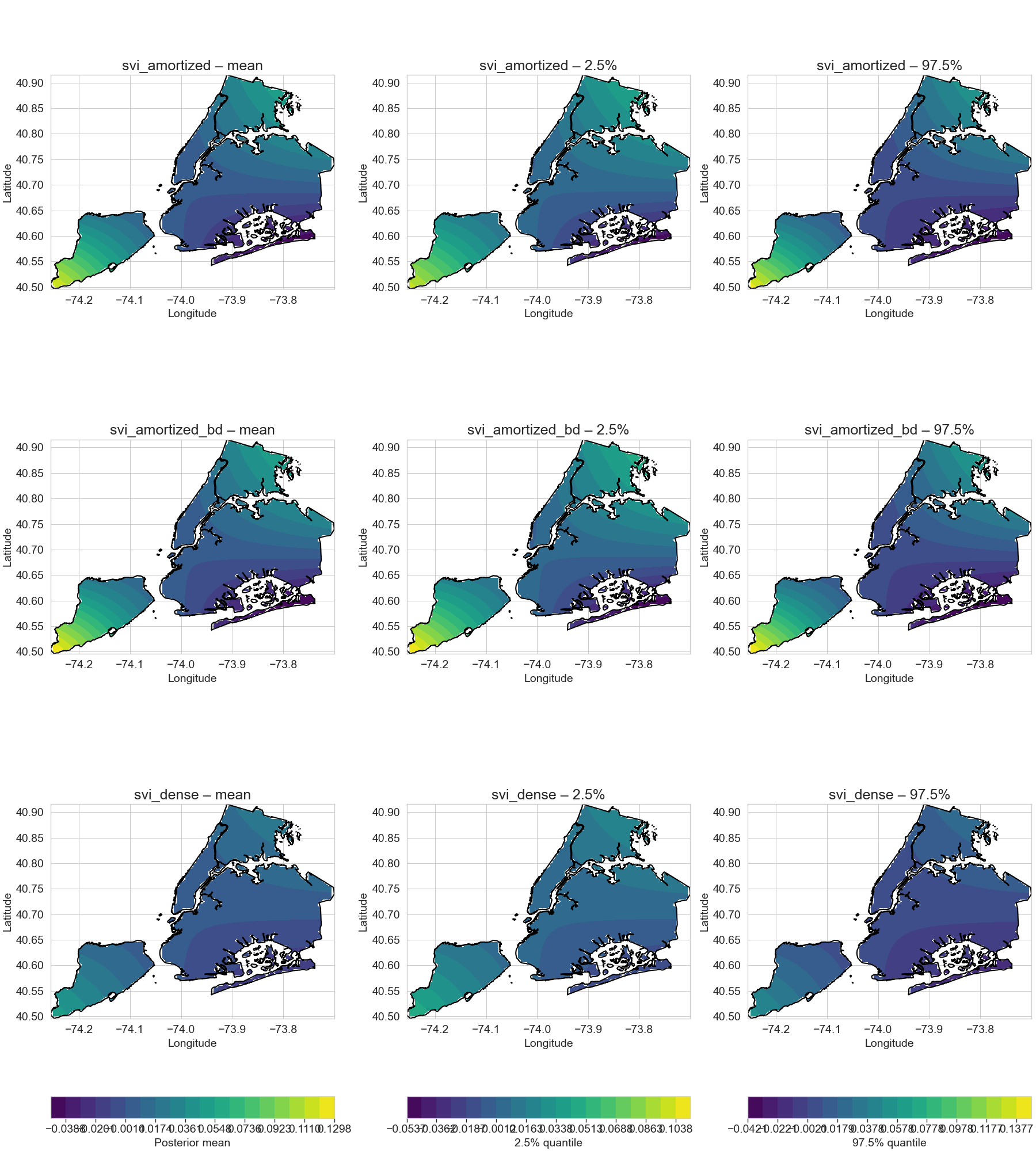}
  \caption{
    Posterior summaries of the spatial tensor-product spline effect in the Poisson model. Each panel shows the smooth contribution of longitude and latitude to the linear predictor, evaluated on a grid within the New York City boundaries. Rows correspond to the three models (\texttt{svi\_amortized}, \texttt{svi\_amortized\_bd}, \texttt{svi\_dense}), while columns display the posterior mean, the $2.5\%$ quantile and the $97.5\%$ quantile. The effect ranges approximately from $-0.30$ to $0.12$, indicating a moderate spatial variation; the \texttt{svi\_dense} specification yields slightly lower estimates overall compared to the two amortized SVI variants.
  }
  \label{fig:ti-poisson-effects}
\end{figure}

Figure~\ref{fig:ti-poisson-effects} displays the posterior summaries of the spatial tensor-product spline term in the Poisson model. The estimated surface exhibits a smooth large-scale spatial gradient over New York City, with values ranging from about $-0.30$ to $0.12$ on the log-intensity scale. This corresponds to an approximate $50\%$ change in expected counts between the lowest and highest locations (from $\exp(-0.30)\approx 0.74$ to $\exp(0.12)\approx 1.13$), suggesting a moderate but non-negligible spatial effect. The three model variants produce very similar spatial patterns and uncertainty bands; the \texttt{svi\_dense} model yields slightly lower estimated effects overall compared to \texttt{svi\_amortized} and \texttt{svi\_amortized\_bd}, consistent with a modest additional shrinkage, but the main spatial features of the effect appear robust to the choice of variational approximation.

\section{Conclusion}\label{section:Conclusion}

In this paper, we introduced an SVI model designed for posterior estimation within the framework of structured additive distributional regression. We proposed two distinct strategies for constructing the variational distribution $q$, leveraging information from the covariates $\X$, potentially the observations $\y$, and the hyperparameters $\bs\lambda$ to expedite the estimation process. The first approach utilizes both the covariates $\X$ and the hyperparameters $\bs\lambda$ to simultaneously learn the location vector and the precision matrix. The second approach addresses the computational complexity of the first by initially assuming independence among the smooth terms $f_{i, p, j}$ and later reintroducing correlation through a set of additional variational parameters. We treat the smoothing parameters $\bs\lambda$ within the variational framework by introducing a dedicated variational approximation for their posterior, which allows joint inference of $\bs\lambda$ and the regression coefficients and accounts for uncertainty in $\bs\lambda$.

We complement these methodological developments with theoretical results establishing sufficient conditions for posterior propriety and, under regularity conditions, consistency and asymptotic normality of the MAP together with a local Gaussian approximation of the posterior. These results provide theoretical support for the Gaussian variational approximations used throughout the proposed framework.

We evaluated our approach against two alternative methods: the widely-used INLA framework~\citep{rue2017bayesian} for Integrated Nested Laplace Approximation and a more advanced version of the variational inference model from~\citep{kleinemeier2023scalable}, which directly estimates the Cholesky factor of the full precision matrix for greater modeling flexibility. In the simulation study detailed in \autoref{section:Simulation}, we assessed the ability of our SVI model to accurately approximate the posterior in structured additive distributional regression. Two scenarios were considered: (i) binary response models, which pose challenges for normal distribution approximations of likelihood terms \cite{albert1993bayesian}, and (ii) gamma response models, where both the location and scale parameters depend on covariates, illustrating a case of distributional regression with a skewed response distribution. Across both tasks, our model variations demonstrated faster convergence to a lower WD compared to the \texttt{dense} model. In the logistic regression case, the posterior estimates for both the regression coefficients $\bs\beta$ and the model parameters $\bs\lambda$ were similar to those produced by MCMC, though INLA exhibited greater variance in terms of WD for $\bs\lambda$. In the gamma regression scenario, our model more accurately replicated the MCMC-estimated posterior distribution of the smoothing parameters than \texttt{svi\_dense}. 

Finally, in \autoref{section:Application}, we demonstrate the scalability of our approach on a large data set of motor vehicle crashes in New York City. Due to the size of the data set, we restrict attention to the three SVI variants and assess convergence using ELBO trajectories over optimization time. In addition, we inspect the resulting estimated covariate effects on the mean parameter to ensure that the fitted effects are sensible. The results show that the proposed amortized schemes reach comparable ELBO values in less computation time than the dense variational baseline, highlighting their computational advantage in a real-data setting.

We plan to extend our framework to handle multivariate responses by employing copula-based constructions \citep{Klein2015Bayesian, klein2016simultaneous, kock2025truly}. This extension would allow us to model dependence between multiple response variables within the structured additive distributional regression framework, thereby broadening the applicability of our approach to more complex data settings.

\bibliographystyle{agsm}

\bibliography{bibliography}

@article{rue2009approximate,
  title={Approximate Bayesian inference for latent Gaussian models by using integrated nested Laplace approximations},
  author={Rue, H{\aa}vard and Martino, Sara and Chopin, Nicolas},
  journal={Journal of the royal statistical society: Series b (statistical methodology)},
  volume={71},
  number={2},
  pages={319--392},
  year={2009},
  publisher={Wiley Online Library}
}

@article{zhang2018advances,
  title={Advances in variational inference},
  author={Zhang, Cheng and B{\"u}tepage, Judith and Kjellstr{\"o}m, Hedvig and Mandt, Stephan},
  journal={IEEE Transactions on Pattern Analysis and Machine Intelligence},
  volume={41},
  number={8},
  pages={2008--2026},
  year={2018},
  publisher={IEEE}
}

@article{kingma2014adam,
  title={Adam: A method for stochastic optimization},
  author={Kingma, Diederik P and Ba, Jimmy},
  journal={arXiv:1412.6980},
  year={2014}
}

@article{kingma2013auto,
  title={Auto-encoding variational {B}ayes},
  author={Kingma, Diederik P and Welling, Max},
  journal={arXiv:1312.6114},
  year={2013}
}

@article{hoffman2013stochastic,
  title={Stochastic variational inference},
  author={Hoffman, Matthew D and Blei, David M and Wang, Chong and Paisley, John},
  journal={J Machine Learning Res},
  volume={14},
  number={1},
  pages={1303--1347},
  year={2013},
  publisher={JMLR. org}
}

@article{blei2017variational,
  title={Variational inference: A review for statisticians},
  author={Blei, David M and Kucukelbir, Alp and McAuliffe, Jon D},
  journal={J Am Stat Assoc},
  volume={112},
  number={518},
  pages={859--877},
  year={2017},
  publisher={Taylor \& Francis}
}

@article{rue2017bayesian,
  title={Bayesian computing with INLA: a review},
  author={Rue, H{\aa}vard and Riebler, Andrea and S{\o}rbye, Sigrunn H and Illian, Janine B and Simpson, Daniel P and Lindgren, Finn K},
  journal={Annual Review of Statistics and Its Application},
  volume={4},
  pages={395--421},
  year={2017},
  publisher={Annual Reviews}
}

@article{kneib2023rage,
  title={Rage against the mean--a review of distributional regression approaches},
  author={Kneib, Thomas and Silbersdorff, Alexander and S{\"a}fken, Benjamin},
  journal={Econometrics and Statistics},
  volume={26},
  pages={99--123},
  year={2023},
  publisher={Elsevier}
}

@misc{liesel,
  author = {Riebl, Hannes and Wiemann, Paul F. V. and Kneib, Thomas},
  title = {Liesel: A Probabilistic Programming Framework for Developing Semi-Parametric Regression Models and Custom Bayesian Inference Algorithms},
  publisher = {arXiv},
  year = {2022},
  url = {https://arxiv.org/abs/2209.10975}
}

@Book{BanCarGel03,
  author = {Banerjee, S. and Carlin, B. P. and Gelfand, A. E.},
  year = {2014},
  title = {Hierarchical Modelling and Analysis for Spatial Data (second edition)},
  publisher = {Chapman \& Hall / CRC}
}

@book{fahkne11,
  title={{B}ayesian Smoothing and Regression for Longitudinal, Spatial and Event History Data},
  author={Fahrmeir, L. and Kneib, T.},
  year={2011},
  publisher={Oxford University Press}
}

@Book{HasTibFri09,
  author = {Hastie, T. J. and Tibshirani, R. J. and Friedman, J.},
  title = {The Elements of Statistical Learning (second edition)},
  publisher = {Springer},
  year = {2016}
}

@article{Klein2015Bayesian,
 author = {Klein, Nadja and Kneib, Thomas and Klasen, Stephan and Lang, Stefan},
 year = {2015},
 title = {Bayesian structured additive distributional regression for multivariate responses},
 pages = {569--591},
 volume = {64},
 issn = {00359254},
 journal = {J Royal Stat Soc C},
 doi = {10.1111/rssc.12090}
}

@article{Klein2015Bayesianb,
 author = {Klein, Nadja and Kneib, Thomas and Lang, Stefan},
 year = {2015},
 title = {Bayesian Generalized Additive Models for Location, Scale, and Shape for Zero-Inflated and Overdispersed Count Data},
 pages = {405--419},
 volume = {110},
 issn = {0162-1459},
 journal = {J Am Stat Assoc},
 doi = {10.1080/01621459.2014.912955}
}

@article{Klein2016Scale,
 author = {Klein, Nadja and Kneib, Thomas},
 year = {2016},
 title = {Scale-Dependent Priors for Variance Parameters in Structured Additive Distributional Regression},
 pages = {1071--1106},
 volume = {11},
 journal = {Bayesian Analysis},
 doi = {10.1214/15-BA983}
}

@article{Klein2016Simultaneous,
 author = {Klein, Nadja and Kneib, Thomas},
 year = {2016},
 title = {Simultaneous inference in structured additive conditional copula regression models: a unifying {B}ayesian approach},
 pages = {841--860},
 volume = {26},
 issn = {0960-3174},
 journal = {Statistics and Computing},
 doi = {10.1007/s11222-015-9573-6}
}

@article{Kneib2013Beyond,
 author = {Kneib, Thomas},
 year = {2013},
 title = {Beyond mean regression},
 pages = {275--303},
 volume = {13},
 issn = {1471-082X},
 journal = {Statistical Modelling},
 doi = {10.1177/1471082X13494159}
}

@Book{rupwan03,
  author =       {Ruppert, D. and Wand, M. P. and Carroll, R. J.},
  title =        {Semiparametric Regression},
  publisher =    {Cambridge University Press},
  year =         {2003}
}

@Book{wood06,
  author =   {Wood, S. N.},
  title =    {Generalized Additive Models: An Introduction with R (second edition)},
  publisher =  {Chapman \& Hall / CRC},
  year =     {2017}
}

@article{fahrmeir2004penalized,
  title={Penalized structured additive regression for space-time data: a Bayesian perspective},
  author={Fahrmeir, Ludwig and Kneib, Thomas and Lang, Stefan},
  journal={Statistica Sinica},
  pages={731--761},
  year={2004},
  publisher={JSTOR}
}

@article{kleinemeier2023scalable,
  title={Scalable estimation for structured additive distributional regression through variational inference},
  author={Kleinemeier, Jana and Klein, Nadja},
  journal={arXiv preprint arXiv:2311.07371},
  year={2023}
}

@article{jerak2006modeling,
  title={Modeling probabilities of patent oppositions in a Bayesian semiparametric regression framework},
  author={Jerak, Alexander and Wagner, Stefan},
  journal={Empirical Economics},
  volume={31},
  pages={513--533},
  year={2006},
  publisher={Springer}
}

@article{pinheiro1996unconstrained,
  title={Unconstrained parametrizations for variance-covariance matrices},
  author={Pinheiro, Jos{\'e} C and Bates, Douglas M},
  journal={Statistics and computing},
  volume={6},
  pages={289--296},
  year={1996},
  publisher={Springer}
}

@article{hoffman2014no,
  title={The No-U-Turn sampler: adaptively setting path lengths in {H}amiltonian {M}onte {C}arlo.},
  author={Hoffman, Matthew D and Gelman, Andrew and others},
  journal={J. Mach. Learn. Res.},
  volume={15},
  number={1},
  pages={1593--1623},
  year={2014}
}

@article{lindgren2015bayesian,
  title={Bayesian spatial modelling with R-INLA},
  author={Lindgren, Finn and Rue, H{\aa}vard},
  journal={Journal of statistical software},
  volume={63},
  number={19},
  year={2015},
  publisher={University of California at Los Angeles}
}

@article{albert1993bayesian,
  title={Bayesian analysis of binary and polychotomous response data},
  author={Albert, James H and Chib, Siddhartha},
  journal={Journal of the American statistical Association},
  volume={88},
  number={422},
  pages={669--679},
  year={1993},
  publisher={Taylor \& Francis}
}

@article{kock2025truly,
  title={Truly multivariate structured additive distributional regression},
  author={Kock, Lucas and Klein, Nadja},
  journal={Journal of Computational and Graphical Statistics},
  volume={34},
  number={4},
  pages={1189--1201},
  year={2025},
  publisher={Taylor \& Francis}
}

@article{Guo2021OnSM,
  title={On Stochastic Moving-Average Estimators for Non-Convex Optimization},
  author={Zhishuai Guo and Yi Xu and Wotao Yin and Rong Jin and Tianbao Yang},
  journal={ArXiv},
  year={2021},
  volume={abs/2104.14840},
  url={https://api.semanticscholar.org/CorpusID:233476430}
}

@article{li2021bayesian,
  title={Bayesian LSTM with stochastic variational inference for estimating model uncertainty in process-based hydrological models},
  author={Li, Dayang and Marshall, Lucy and Liang, Zhongmin and Sharma, Ashish and Zhou, Yan},
  journal={Water Resources Research},
  volume={57},
  number={9},
  pages={e2021WR029772},
  year={2021},
  publisher={Wiley Online Library}
}

@book{villani2008optimal,
  title={Optimal transport: old and new},
  author={Villani, C{\'e}dric and others},
  volume={338},
  year={2008},
  publisher={Springer}
}

@article{flamary2021pot,
  title={Pot: Python optimal transport},
  author={Flamary, R{\'e}mi and Courty, Nicolas and Gramfort, Alexandre and Alaya, Mokhtar Z and Boisbunon, Aur{\'e}lie and Chambon, Stanislas and Chapel, Laetitia and Corenflos, Adrien and Fatras, Kilian and Fournier, Nemo and others},
  journal={Journal of Machine Learning Research},
  volume={22},
  number={78},
  pages={1--8},
  year={2021}
}

\appendix
\section{Appendix}

\subsection{Sketch of the proof of Theorem~1}\label{proof_theorem1}

Sketch of the proof following the investigations in \citet[Section~3.3]{Klein2015Bayesianb} and \citet[Section~3.3]{Klein2016Scale}. For notational convenience, we present the proof for random precision parameters with proper priors. If the precision parameters are fixed, the factors $p(\bs\lambda_p)$ and the integration with respect to $\bs\lambda$ are simply omitted.

\begin{itemize}
    \item Using the reparameterization in (B), for each distributional parameter $p$ we have $\tilde{\bs\eta}_p = \tilde{\U}_p\bs\gamma_p + \tilde{\Z}_p\b_p + \b_p^\varepsilon.$ Since the transformation from $\b_p^\varepsilon$ to $\tilde{\bs\eta}_p$ is one-to-one with unit Jacobian,
    $$
    \b_p^\varepsilon = \tilde{\bs\eta}_p - \tilde{\U}_p\bs\gamma_p - \tilde{\Z}_p\b_p,
    $$
    and hence
    $$
    p(\tilde{\bs\eta}_p\mid\bs\gamma_p,\b_p,\bs\lambda_p) = p_{\b_p^\varepsilon}\left( \tilde{\bs\eta}_p-\tilde{\U}_p\bs\gamma_p-\tilde{\Z}_p\b_p \mid\bs\lambda_p \right).
    $$
    We represent all remaining regression parameters in the vectors $\bs\gamma=(\bs\gamma_1^T,\ldots,\bs\gamma_P^T)^T$, $\b=(\b_1^T,\ldots,\b_P^T)^T$, and $\tilde{\bs\eta}=(\tilde{\bs\eta}_1^T,\ldots,\tilde{\bs\eta}_P^T)^T$ and therefore replace the posterior in \eqref{eq:p(beta, lambda | y, X)} by
    \[
    p(\bs\gamma, \b, \tilde{\bs\eta}, \bs\lambda | \y, \X) \propto \prod_{i=1}^N p( \y_i | \bs\nu_i) \prod_{p=1}^P p(\bs\gamma_p) p(\b_p|\bs\lambda_p) p(\tilde{\bs\eta}_p\mid\bs\gamma_p,\b_p,\bs\lambda_p) p(\bs\lambda_p)
    \]
    We then have that the original posterior is proper if this posterior is proper. 
    \item We now split the data set into observations $i\in\mathcal I^\ast$, for which the density fulfills the integrability condition in (A) and the remaining observations $i\notin\mathcal I^\ast$ for which the density is bounded. Then clearly
    \[
    p(\bs\gamma, \b, \tilde{\bs\eta}, \bs\lambda | \y, \X) \le M^{N-n^\ast} \prod_{i\in\mathcal I^\ast} p( \y_i | \bs\nu_i) \prod_{p=1}^P p(\bs\gamma_p) p(\b_p|\bs\lambda_p) p(\tilde{\bs\eta}_p\mid\bs\gamma_p,\b_p,\bs\lambda_p) p(\bs\lambda_p).
    \]
    \item Noting that $p(\y_i|\bs\nu_i) = p(\y_i|\tilde{\eta}_{i1},\ldots,\tilde{\eta}_{iP})$, we find based on Theorem~3 in \citet[Section~3.3]{Klein2016Scale} (and using assumptions (C) and (D)) that
    \[
    p(\tilde{\bs\eta}|\y,\X) = \int\ldots\int \prod_{i\in\mathcal I^\ast} p(\y_i|\tilde{\eta}_{i1},\ldots,\tilde{\eta}_{iP}) \prod_{p=1}^P p(\bs\gamma_p) p(\b_p|\bs\lambda_p) p(\tilde{\bs\eta}_p|\bs\gamma_p,\b_p,\bs\lambda_p) p(\bs\lambda_p) \,d\bs\gamma\,d\b\,d\bs\lambda
    \]
    is bounded by
    \[
    C \prod_{i\in\mathcal I^\ast} p(\y_i|\tilde{\eta}_{i1},\ldots,\tilde{\eta}_{iP}).
    \]
    \item Therefore, finally, since $p(\y_i|\tilde{\eta}_{i1},\ldots,\tilde{\eta}_{iP})$ is integrable by assumption (A), we find that the overall posterior is proper if assumptions (A) to (D) hold.
\end{itemize}

\clearpage

\subsection{Assumptions for Theorem~2} \label{assumptions_theorem2}

\begin{enumerate}[label=(T2-A\arabic*)]
	\item \label{ass:t2-iid-spec} $(Y_i,\X_i)\stackrel{iid}{\sim}P_{\bs\beta_0}$ for some $\bs\beta_0\in\mathbb{R}^Q$, with $\X_i$ supported on a compact set; the model $p(y \mid \nu(\X, \bs\beta))$ is correctly specified at $\bs\beta_0$.

    \item \label{ass:t2-gaussian-prior} Parameter dimension $Q$ is fixed and so are the smoothing hyperparameters $\bs\lambda$.
    Hence, $p(\bs\beta \mid \bs\lambda) \propto \exp\left(-\frac{1}{2} \bs\beta^\top \K_{\bs\lambda} \bs\beta\right)$ with fixed $\K_{\bs\lambda} \succeq 0$.

    \item \label{ass:t2-identifiable-hessian} The observed posterior information at the MAP is positive definite, with probability tending to 1, $\bm{H}_N(\hat{\bs\beta}_{\bs\lambda}) \succ 0$ for all sufficiently large $N$.

    \item \label{ass:t2-smooth-fisher} For $\bs\beta$ in a neighborhood of $\bs\beta_0$, $\log p(y \mid \nu(\X,\bs\beta))$ is three times continuously differentiable in $\bs\beta$, with score mean zero at $\bs\beta_0$. Moreover, the per-observation expected Fisher information $I(\bs\beta)$ is continuous in a neighborhood of $\bs\beta_0$ and positive definite at $\bs\beta=\bs\beta_0$.

	\item \label{ass:t2-envelope} 
    There exists a neighborhood $\mathcal N$ of $\bs\beta_0$ such that the following local regularity conditions hold.
	For $k=1, 2, 3$, the $k$-th derivatives of the per-observation log-likelihood admit integrable envelopes on $\mathcal N$, i.e., there exist functions $F_k$ with $\mathbb{E}[F_k(\X, Y)] < \infty$ such that
	$$
		\sup_{\bs\beta\in\mathcal N} \left\|
			\nabla_{\bs\beta}^k \log p(Y\mid \nu(\X, \bs\beta))
			\right\| \leq F_k(\X, Y).
	$$
	Moreover, the local uniform LLN holds for the Hessian,
	$$
		\sup_{\bs\beta\in\mathcal N} \left\|
			\frac{1}{N} J_N(\bs\beta) - I(\bs\beta)
		\right\| \xrightarrow{p} 0,
	$$
	and the score at the true parameter satisfies the CLT
	$$
		N^{-1/2} S_N(\bs\beta_0) \xrightarrow{d} \mathcal N\bigl(0, I(\bs\beta_0)\bigr).
	$$

    \item \label{ass:t2-propriety} The assumptions from Theorem~\ref{thm:propriety} hold, such that the posterior is proper.

	\item \label{ass:t2-global-identification} %
    The map
	$$
	M(\bs\beta) = \mathbb E_{\X,Y} \left[\log p(Y\mid\nu(\X,\bs\beta))\right]
	$$
	is continuous and uniquely maximized at $\bs\beta_0$.

	\item \label{ass:t2-coercive} %
	The expected log-likelihood is coercive, that is,
	$$
	\|\bs\beta\| \to \infty \quad \Longrightarrow \quad M(\bs\beta) \to -\infty.
	$$

	\item \label{ass:t2-compact-ulln} For every compact set $C \subseteq \mathbb{R}^Q$,
	$$
	\sup_{\bs\beta \in C}
	\left|
		\frac{1}{N} \ell_N(\bs\beta) - M(\bs\beta)
	\right|
	\xrightarrow{p} 0.
	$$

\item \label{ass:t2-map-tight} %
For every $\epsilon>0$, there exists a compact set $C_\epsilon\subseteq\mathbb R^Q$ with $\bs\beta_0\in C_\epsilon$ such that, for all sufficiently large $N$, with probability greater than $1-\epsilon$, the penalized log-likelihood
$$
L_N(\bs\beta) = \ell_N(\bs\beta) - \frac{1}{2}\bs\beta^\top\K_{\bs\lambda}\bs\beta
$$
admits a unique finite global maximizer
$$
\hat{\bs\beta}_{\bs\lambda} = \arg\max_{\bs\beta\in\mathbb R^Q} L_N(\bs\beta)
$$
satisfying
$$
\hat{\bs\beta}_{\bs\lambda} \in C_\epsilon.
$$

\item \label{ass:t2-integrated-tail-condition} %
For every sufficiently small neighborhood $U$ of $\bs\beta_0$, there exists $\eta_U>0$ such that
$$
\int_{U^c} \exp\left(L_N(\bs\beta)\right) \,d\bs\beta = O_p\left( \exp\left(L_N(\bs\beta_0)-\eta_U N\right)\right),
$$
where
$$
L_N(\bs\beta) = \ell_N(\bs\beta) - \frac{1}{2}\bs\beta^\top\K_{\bs\lambda}\bs\beta.
$$

\end{enumerate}

\clearpage

\subsection{Proof of Theorem~2} \label{proof_theorem2}

\begin{proof}[Proof i.]
Define the average penalized log-likelihood by
$$
	M_N(\bs\beta) = \frac{1}{N} \ell_N(\bs\beta) - \frac{1}{2N}\bs\beta^\top \K_{\bs\lambda}\bs\beta
$$
and let $M$ denote the expected log-likelihood from Assumption~\ref{ass:t2-global-identification}.
Because $\bs\lambda$ is fixed, the penalty contributes $O(N^{-1})$ uniformly on every compact set; hence, by Assumption~\ref{ass:t2-compact-ulln},
$$
	\sup_{\bs\beta\in C} |M_N(\bs\beta) - M(\bs\beta)| \xrightarrow{p} 0
$$
for every compact $C\subset\mathbb R^Q$.
By Assumption~\ref{ass:t2-map-tight}, for every $\delta>0$ there exists a compact set $C_\delta \subseteq \mathbb{R}^Q$ with $\bs\beta_0 \in C_\delta$ such that $P(\hat{\bs\beta}_{\bs\lambda}\in C_\delta) > 1 - \delta$ for all sufficiently large $N$. Since $M$ is continuous and uniquely maximized at $\bs\beta_0$ by Assumption~\ref{ass:t2-global-identification}, the argmax theorem implies that any maximizer of $M_N$ over $C_\delta$ converges in probability to $\bs\beta_0$.

On the event
$$
\left\{\hat{\bs\beta}_{\bs\lambda} \in C_\delta\right\},
$$
the global MAP $\hat{\bs\beta}_{\bs\lambda}$ is the maximizer of $M_N$ over $C_\delta$. Hence, for every $\varepsilon>0$,
$$
\begin{aligned}
P\left(\|\hat{\bs\beta}_{\bs\lambda}-\bs\beta_0\|>\varepsilon\right) &\leq P\left(\|\hat{\bs\beta}_{\bs\lambda}-\bs\beta_0\|>\varepsilon, \hat{\bs\beta}_{\bs\lambda}\in C_\delta \right)
\\
&\quad + P\left(\hat{\bs\beta}_{\bs\lambda}\notin C_\delta \right).
\end{aligned}
$$
The first term converges to zero by the argmax theorem on $C_\delta$, whereas, by Assumption~\ref{ass:t2-map-tight}, the second term is smaller than $\delta$ for all sufficiently large $N$. Therefore,
$$
\limsup_{N\to\infty} P\left(\|\hat{\bs\beta}_{\bs\lambda}-\bs\beta_0\|>\varepsilon\right) \leq \delta.
$$
Since $\delta>0$ is arbitrary,
$$
\hat{\bs\beta}_{\bs\lambda}\xrightarrow{p}\bs\beta_0.
$$

\end{proof}

\begin{proof}[Proof ii.]
Let $S_N(\bs\beta) = \nabla \ell_N(\bs\beta)$ and $J_N(\bs\beta) = -\nabla^2\ell_N(\bs\beta)$. By part~i.,
$$
\hat{\bs\beta}_{\bs\lambda} \xrightarrow{p} \bs\beta_0.
$$
Hence, with probability tending to one, $\hat{\bs\beta}_{\bs\lambda}$ belongs to the neighborhood of $\bs\beta_0$ on which the log-likelihood is differentiable. By Assumption~\ref{ass:t2-map-tight}, the unique finite MAP exists with probability tending to one. Since the parameter space is $\mathbb R^Q$, the first-order condition
$$ 
0 = \nabla L_N(\hat{\bs\beta}_{\bs\lambda}) = S_N(\hat{\bs\beta}_{\bs\lambda}) - \K_{\bs\lambda}\hat{\bs\beta}_{\bs\lambda}
$$
holds with probability tending to one.

Using the fundamental theorem of calculus along the line segment joining $\bs\beta_0$ and $\hat{\bs\beta}_{\bs\lambda}$,
$$ 
S_N(\hat{\bs\beta}_{\bs\lambda}) - S_N(\bs\beta_0)
= \left[\int_0^1 \nabla S_N\left(\bs\beta_0 + t(\hat{\bs\beta}_{\bs\lambda}-\bs\beta_0) \right) \,dt \right] (\hat{\bs\beta}_{\bs\lambda}-\bs\beta_0).$$
Since $\nabla S_N(\bs\beta) = -J_N(\bs\beta)$, define
$$ \bar J_N = \int_0^1 J_N\left(\bs\beta_0 + t(\hat{\bs\beta}_{\bs\lambda}-\bs\beta_0)
\right)\,dt.
$$
Then,
$$
S_N(\hat{\bs\beta}_{\bs\lambda}) = S_N(\bs\beta_0) - \bar J_N(\hat{\bs\beta}_{\bs\lambda}-\bs\beta_0),
$$
and therefore
$$
0 = S_N(\bs\beta_0) - \bar J_N(\hat{\bs\beta}_{\bs\lambda}-\bs\beta_0) - \K_{\bs\lambda}\hat{\bs\beta}_{\bs\lambda}.
$$
Divide by $\sqrt{N}$ and rearrange
$$
\frac{1}{\sqrt{N}}S_N(\bs\beta_0) = \frac{1}{N}\bar J_N\sqrt{N}(\hat{\bs\beta}_{\bs\lambda}-\bs\beta_0) + \frac{1}{\sqrt{N}}\K_{\bs\lambda}\hat{\bs\beta}_{\bs\lambda}
$$
to get
\begin{equation}
\label{eq:t2-diff-dist-rearraged}
\sqrt{N}(\hat{\bs\beta}_{\bs\lambda}-\bs\beta_0) = \left(\frac{1}{N}\bar J_N \right)^{-1} \left(\frac{1}{\sqrt{N}}S_N(\bs\beta_0) - \frac{1}{\sqrt{N}}\K_{\bs\lambda}\hat{\bs\beta}_{\bs\lambda} \right).
\end{equation}
Now, we know
\begin{enumerate}[label=(\alph*)]
	\item By Assumptions~\ref{ass:t2-iid-spec}, \ref{ass:t2-smooth-fisher}, and~\ref{ass:t2-envelope}, the score satisfies a CLT, so
		$$
			N^{-1/2}S_N(\bs\beta_0)
			=
			N^{-1/2}\sum_{i=1}^N s_i(\bs\beta_0)
			\xrightarrow{d} G,
			\qquad
			G \sim \mathcal{N}(0, I(\bs\beta_0)).
		$$

	\item By part~i.,
		$$
		\hat{\bs\beta}_{\bs\lambda} \xrightarrow{p} \bs\beta_0.
		$$
		Hence,
		$$
		\sup_{t\in[0,1]}
		\left\|
		\bs\beta_0+t(\hat{\bs\beta}_{\bs\lambda}-\bs\beta_0)-\bs\beta_0
		\right\|
		\leq
		\|\hat{\bs\beta}_{\bs\lambda}-\bs\beta_0\|
		\xrightarrow{p}0.
		$$
		Thus, with probability tending to one, the entire line segment joining $\bs\beta_0$ and $\hat{\bs\beta}_{\bs\lambda}$ is contained in the neighborhood from Assumption~\ref{ass:t2-envelope}. By the uniform Hessian LLN in Assumption~\ref{ass:t2-envelope} and the continuity of $I(\bs\beta)$ in a neighborhood of $\bs\beta_0$ from Assumption~\ref{ass:t2-smooth-fisher},
		$$
		\frac{1}{N}\bar J_N
		\xrightarrow{p}
		I(\bs\beta_0).
		$$
		Since $I(\bs\beta_0)\succ0$, the inverse exists with probability tending to one and
		$$
		\left(
		\frac{1}{N}\bar J_N
		\right)^{-1}
		\xrightarrow{p}
		I(\bs\beta_0)^{-1}.
		$$

	\item By Assumption~\ref{ass:t2-gaussian-prior}, $\K_{\bs\lambda}$ is fixed. By part~i., $\hat{\bs\beta}_{\bs\lambda} \xrightarrow{p} \bs\beta_0$, so $\hat{\bs\beta}_{\bs\lambda} = O_p(1)$ and therefore
		$$
			N^{-1/2}\K_{\bs\lambda}\hat{\bs\beta}_{\bs\lambda} = o_p(1).
		$$
\end{enumerate}
Combining \eqref{eq:t2-diff-dist-rearraged}, (a)--(c) with Slutsky's theorem gives
$$
\sqrt{N}(\hat{\bs\beta}_{\bs\lambda}-\bs\beta_0)
\xrightarrow{d} W, \qquad W \sim \mathcal{N}(0,I(\bs\beta_0)^{-1}),
$$
which also yields $\|\hat{\bs\beta}_{\bs\lambda}-\bs\beta_0\| = O_p(N^{-1/2})$. Since $I(\bs\beta_0)$ is fixed and positive definite, this implies
$$
\sqrt{N}I(\bs\beta_0)^{1/2}(\hat{\bs\beta}_{\bs\lambda}-\bs\beta_0) \xrightarrow{d} Z, \qquad Z \sim \mathcal{N}(0,I_Q).
$$
Moreover, by part~i.\ and the continuity of $I(\bs\beta)$ in a neighborhood of $\bs\beta_0$ from Assumption~\ref{ass:t2-smooth-fisher}, the continuous mapping theorem gives
$$
I(\hat{\bs\beta}_{\bs\lambda})^{1/2}I(\bs\beta_0)^{-1/2}
\xrightarrow{p} I_Q.
$$
Finally, write
$$
\sqrt{N}I(\hat{\bs\beta}_{\bs\lambda})^{1/2}(\hat{\bs\beta}_{\bs\lambda}-\bs\beta_0)
=
\left[
I(\hat{\bs\beta}_{\bs\lambda})^{1/2}I(\bs\beta_0)^{-1/2}
\right]
\left[
\sqrt{N}I(\bs\beta_0)^{1/2}(\hat{\bs\beta}_{\bs\lambda}-\bs\beta_0)
\right].
$$
So,
$$
\sqrt{N}I(\hat{\bs\beta}_{\bs\lambda})^{1/2}(\hat{\bs\beta}_{\bs\lambda}-\bs\beta_0)
\xrightarrow{d} Z,
\qquad
Z \sim \mathcal{N}(0,I_Q),
$$
by Slutsky's theorem.
\end{proof}

\begin{proof}[Proof iii.]
Let
$$
L_N(\bs\beta) = \ell_N(\bs\beta) - \frac{1}{2}\bs\beta^\top \K_{\bs\lambda}\bs\beta.
$$
By part~i.\ and Assumption~\ref{ass:t2-map-tight}, with probability tending to one the finite MAP lies in the neighborhood of $\bs\beta_0$ where $L_N$ is differentiable. Since the parameter space is $\mathbb R^Q$,
$$
\nabla L_N(\hat{\bs\beta}_{\bs\lambda})=0.
$$
For brevity, write
$$
H_N = H_N(\hat{\bs\beta}_{\bs\lambda}).
$$
A second-order Taylor expansion of $L_N$ at $\hat{\bs\beta}_{\bs\lambda}$ with third-order remainder gives
$$
L_N(\bs\beta) = L_N(\hat{\bs\beta}_{\bs\lambda}) -\frac{1}{2}(\bs\beta-\hat{\bs\beta}_{\bs\lambda})^\top H_N (\bs\beta-\hat{\bs\beta}_{\bs\lambda}) + R_N(\bs\beta),
$$
with remainder satisfying

\begin{align*}
|R_N(\bs\beta)| &\leq \frac{1}{6}\sup_{t\in[0,1]} \left\| \nabla^3L_N \left(\hat{\bs\beta}_{\bs\lambda}+t(\bs\beta-\hat{\bs\beta}_{\bs\lambda})\right) \right\| \|\bs\beta-\hat{\bs\beta}_{\bs\lambda}\|^3
\\
&= \frac{1}{6} \sup_{t\in[0,1]}\left\| \nabla^3\ell_N \left( \hat{\bs\beta}_{\bs\lambda}+t(\bs\beta-\hat{\bs\beta}_{\bs\lambda})\right) \right\|\|\bs\beta-\hat{\bs\beta}_{\bs\lambda}\|^3.
\end{align*}

	By part~i.\ and since $r_N \to 0$, $B_N$ is contained in the neighborhood $\mathcal N$ of $\bs\beta_0$ from Assumption~\ref{ass:t2-envelope} with probability tending to one.
	By Assumption~\ref{ass:t2-envelope}, there exists a function $F_3(\X, Y)$ with finite expectation such that

$$
\left\| \nabla_{\bs\beta}^3 \log p(Y\mid\nu(\X,\bs\beta)) \right\| \leq F_3(\X,Y).
$$

for all $\bs\beta$ in a sufficiently small neighborhood $\mathcal N$ of
$\bs\beta_0$. Thus,
$$
\sup_{\bs\beta\in\mathcal N} \|\nabla^3\ell_N(\bs\beta)\| \leq \sum_{i=1}^N F_3(\X_i,Y_i) = O_p(N).
$$
Therefore, uniformly for $\bs\beta\in B_N$,
$$
|R_N(\bs\beta)| = O_p(Nr_N^3).
$$
Since
$$
r_N = c\sqrt{\frac{\log N}{N}},
$$
we have
$$
Nr_N^3 = c^3\frac{(\log N)^{3/2}}{N^{1/2}} \to 0.
$$
Hence,
$$
\sup_{\bs\beta\in B_N} |R_N(\bs\beta)| = O_p\left(\frac{(\log N)^{3/2}}{N^{1/2}} \right) = o_p(1).
$$
\end{proof}

\begin{proof}[Proof iv.]

Recall $L_N(\bs\beta) = \ell_N(\bs\beta) -\frac12\bs\beta^\top \K_{\bs\lambda}\bs\beta$,
and let
$$
r_N=c\sqrt{\frac{\log N}{N}}, \qquad B_N= \left\{\bs\beta\in\mathbb R^Q: \Vert\bs\beta-\hat{\bs\beta}_{\bs\lambda}\Vert\le r_N \right\}.
$$

Fix a radius $\delta > 0$ small enough that the closed ball
$$
    U = \{\bs\beta \in \mathbb R^Q : \Vert \bs\beta - \bs\beta_0 \Vert \leq \delta\}
$$
is contained in the neighborhood $\mathcal N$ from Assumption~\ref{ass:t2-envelope} and that Assumption~\ref{ass:t2-integrated-tail-condition} applies to $U$. Lower bounding $\int_{\mathbb R^Q} e^{L_N(\bs\beta)} d\bs\beta$ by the integral over $B_N$ and splitting $B_N^c \subseteq (U \cap B_N^c) \cup U^c$ gives
\begin{align*}
\Pi_N(B_N^c) &= \frac{
		\int_{B_N^c}e^{L_N(\bs\beta)}d\bs\beta
	}{
		\int_{\mathbb R^Q}e^{L_N(\bs\beta)} d\bs\beta
	}\\   
&\leq \frac{\int_{B_N^c} e^{L_N(\bs\beta)}d\bs\beta}{\int_{B_N} e^{L_N(\bs\beta)} d\bs\beta}\\
&\leq
	\underbrace{
	\frac{\int_{U\cap B_N^c} e^{L_N(\bs\beta)} d\bs\beta}{\int_{B_N} e^{L_N(\bs\beta)} d\bs\beta}
	}_{=: T_{N,\text{near}}}
	+
	\underbrace{
		\frac{\int_{U^c} e^{L_N(\bs\beta)} d\bs\beta}{\int_{B_N} e^{L_N(\bs\beta)}d\bs\beta}
	}_{=: T_{N,\text{far}}}.
\end{align*}
We show $T_{N,\text{near}} \xrightarrow{p} 0$ and $T_{N,\text{far}} \xrightarrow{p} 0$ separately.

\paragraph{Step 1: A high-probability event on $U$.}
We construct a high-probability event on which $L_N$ has uniformly bounded quadratic curvature on $U$.
We show that there exist constants $0<\underline{\kappa}\leq\overline{\kappa}<\infty$ (depending on $\bs\beta_0$ and $\delta$ but not on $N$) such that, with probability tending to one,
\begin{align}\label{eq:t2-curvature}
	\underline{\kappa}N I_Q \preceq H_N(\bs\beta) \preceq \overline{\kappa}N I_Q
\qquad\text{for all } \bs\beta \in U.
\end{align}

By Assumption~\ref{ass:t2-envelope}, the uniform LLN for the Hessian gives
$$
    \sup_{\bs\beta\in U}\left\Vert \frac{1}{N}J_N(\bs\beta) - I(\bs\beta) \right\Vert \xrightarrow{p} 0.
$$
By Assumption~\ref{ass:t2-smooth-fisher}, $I(\bs\beta_0) \succ 0$ and $\bs\beta\mapsto I(\bs\beta)$ is continuous near $\bs\beta_0$. Hence, for $\delta$ small enough, there exist constants $0 < \underline{\kappa}_0 \leq \overline{\kappa}_0 < \infty$ with
$$
	2 \underline{\kappa}_0 \leq \inf_{\bs\beta\in U}\lambda_{\min} \big(I(\bs\beta)\big)
\quad\text{and}\quad
	\sup_{\bs\beta\in U}\lambda_{\max} \big(I(\bs\beta)\big) \leq \overline{\kappa}_0/2.
$$

The uniform LLN above then implies that with probability tending to one,
$$
\underline{\kappa}_0 I_Q \preceq \frac{1}{N}J_N(\bs\beta) \preceq \overline{\kappa}_0 I_Q \qquad\text{for all }\bs\beta \in U.
$$
Since $H_N(\bs\beta)=J_N(\bs\beta)+\K_{\bs\lambda}$ with
$\K_{\bs\lambda}\succeq 0$ fixed by Assumption (T2-A2),
$$
\underline{\kappa}_0NI_Q \preceq H_N(\bs\beta)
\preceq \overline{\kappa}_0NI_Q+\K_{\bs\lambda}
\preceq 2\overline{\kappa}_0NI_Q
$$
for all $\bs\beta\in U$ and all $N$ large enough that $\K_{\bs\lambda}\preceq \overline{\kappa}_0NI_Q$, which holds since $\K_{\bs\lambda}$ is fixed and $N\to\infty$. Hence, \eqref{eq:t2-curvature} holds with $\underline{\kappa}=\underline{\kappa}_0$ and $\overline{\kappa}=2\overline{\kappa}_0$.

Define
$$
c_0 := \sqrt{\frac{2Q}{\underline{\kappa}}}.
$$
For the remainder of the proof, fix any $c>c_0$.

Define the event
$$
E_N = \left\{\Vert\hat{\bs\beta}_{\bs\lambda}-\bs\beta_0\Vert \leq \frac{\delta}{2}\right\}\cap\left\{\eqref{eq:t2-curvature}\text{ holds}\right\}.
$$
By part i.\ and the argument above, $P(E_N)\to 1$. Throughout Steps 2 and 3, we work on $E_N$.

\paragraph{Step 2: Control of $T_{N,\text{near}}$.}
Fix $\rho \in (0, 1)$ to be chosen below and define the smaller ball
$$
B_N(\rho) = \{\bs\beta : \Vert\bs\beta - \hat{\bs\beta}_{\bs\lambda}\Vert \leq \rho r_N\} \subseteq B_N.
$$
On $E_N$, for $N$ large enough that $\rho r_N \leq \delta/2$, we have $B_N(\rho) \subseteq U$ since for any $\bs\beta \in B_N(\rho)$,
$$
\Vert\bs\beta - \bs\beta_0\Vert \leq \Vert\bs\beta - \hat{\bs\beta}_{\bs\lambda}\Vert + \Vert\hat{\bs\beta}_{\bs\lambda} - \bs\beta_0\Vert \leq \rho r_N + \delta/2 \leq \delta.
$$

By part~i.\ and Assumption~\ref{ass:t2-map-tight}, with probability tending to one the finite MAP lies in the neighborhood of $\bs\beta_0$ where $L_N$ is differentiable. Since the parameter space is
$\mathbb R^Q$,
$$
\nabla L_N(\hat{\bs\beta}_{\bs\lambda})=0.
$$
Taylor's theorem with integral remainder, applied to $L_N$ at $\hat{\bs\beta}_{\bs\lambda}$, therefore gives
$$
L_N(\bs\beta)-L_N(\hat{\bs\beta}_{\bs\lambda}) = -\int_0^1(1-t) (\bs\beta-\hat{\bs\beta}_{\bs\lambda})^\top H_N\left(\hat{\bs\beta}_{\bs\lambda}+t(\bs\beta-\hat{\bs\beta}_{\bs\lambda}) \right) (\bs\beta-\hat{\bs\beta}_{\bs\lambda}) \,dt.
$$

For $\bs\beta \in U$, by convexity of $U$ the entire segment $\hat{\bs\beta}_{\bs\lambda} + t(\bs\beta - \hat{\bs\beta}_{\bs\lambda})$ lies in $U$ for $t \in [0,1]$. Applying~\eqref{eq:t2-curvature} pointwise inside the integrand and using $\int_0^1 (1-t) dt = 1/2$ yields
\begin{equation}\label{eq:t2-quadratic}
-\tfrac{1}{2} \overline{\kappa}N \Vert \bs\beta - \hat{\bs\beta}_{\bs\lambda} \Vert^2
	\leq L_N(\bs\beta) - L_N(\hat{\bs\beta}_{\bs\lambda})
	\leq -\tfrac{1}{2} \underline{\kappa}N \Vert \bs\beta - \hat{\bs\beta}_{\bs\lambda} \Vert^2
\qquad \text{for all }\bs\beta\in U.
\end{equation}

Now,
\begin{itemize}
	\item for the numerator of $T_{N,\text{near}}$, every $\bs\beta \in U \cap B_N^c$ satisfies $\Vert\bs\beta - \hat{\bs\beta}_{\bs\lambda}\Vert \geq r_N$, so the right inequality gives
$$
	L_N(\bs\beta) \leq
		L_N(\hat{\bs\beta}_{\bs\lambda}) - \tfrac{1}{2} \underline{\kappa} N r_N^2
		= L_N(\hat{\bs\beta}_{\bs\lambda}) - \tfrac{\underline{\kappa} c^2}{2} \log N,
	$$
	and therefore
$$
	\int_{U \cap B_N^c} e^{L_N(\bs\beta)} d\bs\beta
	    \leq e^{L_N(\hat{\bs\beta}_{\bs\lambda})} N^{-\underline{\kappa} c^2 / 2} \mathrm{Vol}(U).
	$$

	\item for the denominator, every $\bs\beta\in B_N(\rho)$ satisfies $\Vert\bs\beta - \hat{\bs\beta}_{\bs\lambda}\Vert \leq \rho r_N$, so the left inequality of~\eqref{eq:t2-quadratic} gives
$$
	L_N(\bs\beta)
	\geq L_N(\hat{\bs\beta}_{\bs\lambda}) - \tfrac{1}{2} \overline{\kappa} N (\rho r_N)^2
	= L_N(\hat{\bs\beta}_{\bs\lambda}) - \tfrac{\overline{\kappa} \rho^2 c^2}{2}\log N,
$$
	and therefore
	\begin{equation}\label{eq:t2-denom-bound}
	\int_{B_N} e^{L_N(\bs\beta)} d\bs\beta
	    \geq \int_{B_N(\rho)} e^{L_N(\bs\beta)} d\bs\beta
	    \geq e^{L_N(\hat{\bs\beta}_{\bs\lambda})} N^{-\overline{\kappa}\rho^2 c^2 / 2} \mathrm{Vol}(B_N(\rho)).
	\end{equation}
\end{itemize}

Taking the ratio gives, on $E_N$,
$$
T_{N,\text{near}} \leq \frac{\mathrm{Vol}(U)}{\mathrm{Vol}(B_N(\rho))} N^{-(\underline{\kappa} - \overline{\kappa}\rho^2) c^2 / 2}.
$$
Since $B_N(\rho)$ is a $Q$-dimensional ball of radius $\rho r_N$, $\mathrm{Vol}(B_N(\rho)) = C_Q(\rho r_N)^Q$ with $C_Q = \pi^{Q/2}/\Gamma(Q/2 + 1)$, so
$$
\frac{\mathrm{Vol}(U)}{\mathrm{Vol}(B_N(\rho))}
    = \frac{\mathrm{Vol}(U)}{C_Q \rho^Q} r_N^{-Q}
\asymp r_N^{-Q}
= \left(\frac{N}{\log N}\right)^{Q/2}.
$$
Choose $\rho>0$ such that
$$
\rho^2 < \frac{\underline{\kappa}}{2\overline{\kappa}},
$$
so that
$$
\underline{\kappa}-\overline{\kappa}\rho^2 > \frac{\underline{\kappa}}{2}.
$$
Then, on $E_N$,
$$
T_{N,\mathrm{near}} \lesssim \left(\frac{N}{\log N}\right)^{Q/2} N^{-\underline{\kappa}c^2/4} = \frac{N^{Q/2-\underline{\kappa}c^2/4}}{(\log N)^{Q/2}}.
$$
Since $c>c_0$ and
$$
c_0^2=\frac{2Q}{\underline{\kappa}},
$$
we have
$$
c^2>\frac{2Q}{\underline{\kappa}},
$$
and therefore
$$
\frac{Q}{2}-\frac{\underline{\kappa}c^2}{4}<0.
$$
It follows that
$$
T_{N,\mathrm{near}}\to 0
$$
on $E_N$. Since $P(E_N)\to 1$, we conclude that
$$
T_{N,\mathrm{near}}\xrightarrow{p}0.
$$

\paragraph{Step 3: Control of $T_{N,\text{far}}$.}
The neighborhood $U$ was chosen small enough such that Assumption~\ref{ass:t2-integrated-tail-condition} holds, which gives $\eta_U > 0$ such that
\begin{equation}\label{eq:t2-tail-numerator}
	\int_{U^c} e^{L_N(\bs\beta)} d\bs\beta = O_p\left( e^{L_N(\bs\beta_0) - \eta_U N} \right).
\end{equation}
For the denominator, the lower bound~\eqref{eq:t2-denom-bound} from above, combined with the property $L_N(\hat{\bs\beta}_{\bs\lambda}) \geq L_N(\bs\beta_0)$, yields on $E_N$
\begin{equation}\label{eq:t2-denom-bound-beta0}
	\int_{B_N} e^{L_N(\bs\beta)} d\bs\beta
		\geq e^{L_N(\bs\beta_0)} N^{-\overline{\kappa} \rho^2 c^2 / 2} \mathrm{Vol}(B_N(\rho)).
\end{equation}
Combining~\eqref{eq:t2-tail-numerator} and \eqref{eq:t2-denom-bound-beta0}, and using the fact that
$$
\frac{\int_{U^c}e^{L_N(\bs\beta)}\,d\bs\beta}{e^{L_N(\bs\beta_0)-\eta_U N}} = O_p(1),
$$
we obtain
$$
T_{N,\mathrm{far}} = O_p\left(e^{-\eta_U N}\frac{N^{\overline{\kappa}\rho^2c^2/2}}{\mathrm{Vol}(B_N(\rho))}\right).
$$
Using
$$
\mathrm{Vol}(B_N(\rho))^{-1} \asymp \left(\frac{N}{\log N}\right)^{Q/2},
$$
it follows that
$$
T_{N,\mathrm{far}} = O_p\left(e^{-\eta_U N} N^{\overline{\kappa}\rho^2c^2/2+Q/2} (\log N)^{-Q/2} \right) = o_p(1),
$$
since the exponential factor $e^{-\eta_U N}$ dominates every polynomial in $N$. Hence,
$$
T_{N,\mathrm{far}}\xrightarrow{p}0.
$$

\paragraph{Combine.} Combining Steps 2 and 3,
$$
\Pi_N(B_N^c) \leq T_{N,\text{near}} + T_{N,\text{far}} \xrightarrow{p} 0.
$$
\end{proof}

\begin{proof}[Proof v.]
For brevity, write $H_N = H_N(\hat{\bs\beta}_{\bs\lambda})$. Write the third-order remainder from part~iii.\ as
$$
R_N(\bs\beta)
=
L_N(\bs\beta)-L_N(\hat{\bs\beta}_{\bs\lambda})
+
\frac{1}{2}
(\bs\beta-\hat{\bs\beta}_{\bs\lambda})^\top
H_N
(\bs\beta-\hat{\bs\beta}_{\bs\lambda}).
$$
Let
$$
\varepsilon_N
=
\sup_{\bs\beta\in B_N}|R_N(\bs\beta)|.
$$
By part~iii.,
$$
\varepsilon_N\xrightarrow{p}0.
$$

Define the unnormalized densities
$$
v_N(\bs\beta)
=
\exp\left(
-\frac{1}{2}
(\bs\beta-\hat{\bs\beta}_{\bs\lambda})^\top
H_N
(\bs\beta-\hat{\bs\beta}_{\bs\lambda})
\right)
\indicator{\bs\beta\in B_N}
$$
and
$$
u_N(\bs\beta)
=
\exp(L_N(\bs\beta))
\indicator{\bs\beta\in B_N}.
$$
By the definition of $R_N(\bs\beta)$,
$$
u_N(\bs\beta)
=
\exp(L_N(\hat{\bs\beta}_{\bs\lambda}))
v_N(\bs\beta)
\exp(R_N(\bs\beta)).
$$

Let
$$
Z_N^v = \int_{B_N}v_N(\bs\beta)\,d\bs\beta.
$$
Let $\Phi_N$ denote the Gaussian probability measure with mean
$\hat{\bs\beta}_{\bs\lambda}$ and covariance $H_N^{-1}$, and define its
conditional version on $B_N$ by
$$
\Phi_N^B(A)
=
\Phi_N(A\mid B_N)
=
\frac{\Phi_N(A\cap B_N)}{\Phi_N(B_N)}.
$$
Similarly, define the posterior measure conditioned on $B_N$ by
$$
\Pi_N^B(A)
=
\Pi_N(A\mid B_N)
=
\frac{\Pi_N(A\cap B_N)}{\Pi_N(B_N)}.
$$
Equivalently,
$$
\Pi_N^B(A)
=
\frac{
\displaystyle
\int_{A\cap B_N}e^{L_N(\bs\beta)}\,d\bs\beta
}{
\displaystyle
\int_{B_N}e^{L_N(\bs\beta)}\,d\bs\beta
}.
$$

For $\bs\beta\in B_N$, using
$$
\exp(L_N(\bs\beta))
=
\exp(L_N(\hat{\bs\beta}_{\bs\lambda}))
v_N(\bs\beta)
\exp(R_N(\bs\beta)),
$$
we obtain
$$
\begin{aligned}
\int_{A\cap B_N}\exp(L_N(\bs\beta))\,d\bs\beta
&=
\exp(L_N(\hat{\bs\beta}_{\bs\lambda}))
\int_{A\cap B_N}
v_N(\bs\beta)\exp(R_N(\bs\beta))\,d\bs\beta
\\
&=
\exp(L_N(\hat{\bs\beta}_{\bs\lambda}))
Z_N^v
\mathbb E_{\Phi_N^B}
\left[
\exp(R_N(\bs\beta))
\indicator{\bs\beta\in A}
\right].
\end{aligned}
$$
Similarly,
$$
\int_{B_N}\exp(L_N(\bs\beta))\,d\bs\beta
=
\exp(L_N(\hat{\bs\beta}_{\bs\lambda}))
Z_N^v
\mathbb E_{\Phi_N^B}
\left[
\exp(R_N(\bs\beta))
\right].
$$
Therefore,
$$
\Pi_N^B(A)
=
\frac{
\mathbb E_{\Phi_N^B}
\left[
\exp(R_N(\bs\beta))
\indicator{\bs\beta\in A}
\right]
}{
\mathbb E_{\Phi_N^B}
\left[
\exp(R_N(\bs\beta))
\right]
}.
$$
Thus, the Radon--Nikodym derivative of $\Pi_N^B$ with respect to
$\Phi_N^B$ is
$$
h_N(\bs\beta)
:=
\frac{
\exp(R_N(\bs\beta))
}{
\mathbb E_{\Phi_N^B}
\left[
\exp(R_N(\bs\beta))
\right]
}.
$$

For any $A\in\mathcal B(\mathbb R^Q)$,
$$
\Pi_N^B(A)-\Phi_N^B(A)
=
\int_A
\left(
h_N(\bs\beta)-1
\right)
d\Phi_N^B(\bs\beta).
$$
Hence,
$$
\sup_{A\in\mathcal B(\mathbb R^Q)}
\left|
\Pi_N^B(A)-\Phi_N^B(A)
\right|
\leq
\sup_{\bs\beta\in B_N}
|h_N(\bs\beta)-1|.
$$

By part~iii., on $B_N$ we have
$$
|R_N(\bs\beta)|\leq\varepsilon_N,
$$
and therefore
$$
\exp(-\varepsilon_N)
\leq
\exp(R_N(\bs\beta))
\leq
\exp(\varepsilon_N).
$$
Since $\Phi_N^B$ is a probability measure, the same bounds hold for
the expectation:
$$
\exp(-\varepsilon_N)
\leq
\mathbb E_{\Phi_N^B}
\left[
\exp(R_N(\bs\beta))
\right]
\leq
\exp(\varepsilon_N).
$$
Consequently,
$$
\exp(-2\varepsilon_N)
\leq
h_N(\bs\beta)
\leq
\exp(2\varepsilon_N),
$$
and hence
$$
\sup_{\bs\beta\in B_N}
|h_N(\bs\beta)-1|
\leq
\exp(2\varepsilon_N)-1.
$$
Since
$$
\varepsilon_N\xrightarrow{p}0,
$$
we obtain
$$
\sup_{A\in\mathcal B(\mathbb R^Q)}
\left|
\Pi_N^B(A)-\Phi_N^B(A)
\right|
\xrightarrow{p}0.
$$

It remains to relate the conditional probability measures $\Pi_N^B$ and $\Phi_N^B$ to the unconditional measures.

For every $A\in\mathcal B(\mathbb R^Q)$,
$$
\Pi_N(A\cap B_N) = \Pi_N(B_N)\Pi_N^B(A)
$$
and
$$
\Phi_N(A\cap B_N) = \Phi_N(B_N)\Phi_N^B(A).
$$
Therefore,
$$
\begin{aligned}
&
\left| \Pi_N(A\cap B_N) - \Phi_N(A\cap B_N) \right|
\\
&\qquad
= \left| \Pi_N(B_N)\Pi_N^B(A) - \Phi_N(B_N)\Phi_N^B(A) \right|
\\
&\qquad \leq \Pi_N(B_N) \left| \Pi_N^B(A)-\Phi_N^B(A) \right| + \Phi_N^B(A) \left| \Pi_N(B_N)-\Phi_N(B_N) \right|
\\
&\qquad
\leq \left| \Pi_N^B(A)-\Phi_N^B(A) \right| + \left| \Pi_N(B_N)-\Phi_N(B_N) \right|.
\end{aligned}
$$
Moreover,
$$
\left|\Pi_N(B_N)-\Phi_N(B_N)\right| = \left| \Pi_N(B_N^c)-\Phi_N(B_N^c) \right| \leq \Pi_N(B_N^c)+\Phi_N(B_N^c).
$$
Part~iv.\ gives
$$
\Pi_N(B_N^c)\xrightarrow{p}0.
$$

It remains to show that
$$
\Phi_N(B_N^c)\xrightarrow{p}0.
$$
Let $\underline{\kappa}>0$ be the constant from the curvature bound \eqref{eq:t2-curvature}. On the event where
$$
H_N(\hat{\bs\beta}_{\bs\lambda})\succeq \underline{\kappa}NI_Q,
$$
for every $\bs\beta\in B_N^c$ we have
$$
\begin{aligned}
(\bs\beta-\hat{\bs\beta}_{\bs\lambda})^\top H_N(\hat{\bs\beta}_{\bs\lambda}) (\bs\beta-\hat{\bs\beta}_{\bs\lambda})
&\geq \underline{\kappa}N \|\bs\beta-\hat{\bs\beta}_{\bs\lambda}\|^2
\\
&\geq \underline{\kappa}Nr_N^2
\\
&=
\underline{\kappa}c^2\log N.
\end{aligned}
$$

Under $\Phi_N$, define
$$
Z = H_N(\hat{\bs\beta}_{\bs\lambda})^{1/2} (\bs\beta-\hat{\bs\beta}_{\bs\lambda}).
$$
Then
$$
Z\sim\mathcal N(0,I_Q),
$$
and hence
$$
\|Z\|^2\sim\chi_Q^2.
$$
Therefore, on the curvature event,
$$
\Phi_N(B_N^c) \leq P\left(\|Z\|^2\geq \underline{\kappa}c^2\log N\right).
$$
By Markov's inequality,
$$
P\left(\|Z\|^2\geq \underline{\kappa}c^2\log N \right) \leq \frac{\mathbb E\|Z\|^2}{\underline{\kappa}c^2\log N} = \frac{Q}{\underline{\kappa}c^2\log N} \to0.
$$
Since the curvature event has probability tending to one, it follows that
$$
\Phi_N(B_N^c)\xrightarrow{p}0.
$$

Combining this with part~iv.\ and the conditional result established above gives
$$
\begin{aligned}
&
\sup_{A\in\mathcal B(\mathbb R^Q)} \left| \Pi_N(A\cap B_N) - \Phi_N(A\cap B_N) \right|
\\ &\qquad \leq \sup_{A\in\mathcal B(\mathbb R^Q)} \left| \Pi_N^B(A)-\Phi_N^B(A) \right| + \Pi_N(B_N^c) + \Phi_N(B_N^c) \xrightarrow{p}0.
\end{aligned}
$$
This proves the local Gaussianity statement in part~v.

\end{proof}

\clearpage

\subsection{Identifiability}\label{subsec:additive models}

In additive regression models, the additive effects have to be constrained to ensure
identifiability since
\begin{align*}
    f_{p, 1}(\x_i, \bs\beta_{p, 1}) + f_{p, 2}(\x_i, \bs\beta_{p, 2}) = f_{p, 1}(\x_i, \bs\beta_{p, 1}) + c + f_{p, 2}(\x_i, \bs\beta_{p, 2}) - c
\end{align*}
This is usually achieved by
\begin{align*}
    \sum_{i}^{N} f^{\text{new}}_{p, j}(x_{i, j}, \bs\beta_{p, j}) = 0,
\end{align*}
with
\begin{align*}
    f_{p, j}(x_{i, j}, \bs\beta_{p, j}) = f^{\text{new}}_{p, j}(x_{i, j}, \bs\beta_{p, j}) + \beta_{0, p, j}.
\end{align*}
To do so, we calculate the vector $\C_{p, j}$ of column means of $\X_{p, j}$ and we build the QR-decomposition of
\begin{align*}
    \C_{p, j} = \Q_{p, j}\R_{p, j},
\end{align*}
where $\Q_{p, j}$ is an orthogonal matrix and 
\begin{align*}
    \R_{p, j} = (\tilde R_{p, j}, 0, \dots, 0)^\T,
\end{align*}
where $\tilde R_{p, j}$ is the upper triangular matrix of dimension $1 \times 1$. Then, we split 
\begin{align*}
    \Q_{p, j} = \left[ \mathbf{Z}_{p, j}^{(a)}, \mathbf{Z}_{p, j}^{(b)} \right],
\end{align*}
where $\mathbf{Z}_{p, j}^{(a)}$ is a $Q_{p, j} \times 1$ column vector. Finally, we write the new design as
\begin{align*}
    \X_{p, j}^{\text{new}} = \X_{p, j}\mathbf{Z}_{p, j}^{(b)}
\end{align*}
and the new penalty matrix as:
\begin{align*}
    \K_{p, j}^{\text{new}}(\lambda_{p, j}) = \left(\mathbf{Z}_{p, j}^{(b)}\right)^\T \K_{p, j} \mathbf{Z}_{p, j}^{(b)}.
\end{align*}
The design matrix for one parameter $p$ can be written as:
\begin{align}
    \X_p = \left[\mathbf{1}, \X_{p, 1}^{\text{new}}, \dots, \X_{p, J_p}^{\text{new}} \right] \in \mathbb{R}^{N \times Q_p}
    \label{eq:X_p}
\end{align}
where $Q_p \coloneqq (Q_{p, 1} + \dots + Q_{p, J_p})$ and the penalty matrix as:
\begin{align*}
    \K_p(\bs\lambda_p) = 
    \begin{pmatrix}
    0 & \dots & 0\\
    0 & \lambda^2_{p, 1}\K_{p, 1}^{\text{new}} & \\
    \vdots & \ddots & \\
    0 & \dots & \lambda^2_{p, J_p}\K_{p, J_p}^{\text{new}}\\
    \end{pmatrix}.
\end{align*}

\clearpage

\subsection{Mini-batching: Unbiased Estimators and ELBO Bias}
\label{app:mini-batching}

This appendix provides the full derivation of the mini-batch estimators used in our
amortized variational inference scheme. In the main text, we introduced scaled
mini-batch estimators for the precision matrix and the natural mean, which allow the
variational distribution to be constructed from a random subset of the data while
keeping the stochastic gradients unbiased. Here, we formally establish the
unbiasedness of these estimators and analyze the bias that arises when
mini-batching is used to approximate the ELBO itself. While this bias is typically
negligible in practice, understanding its origin clarifies the behaviour of stochastic
optimization in amortized variational inference.

\begin{theorem}[Unbiasedness of the Precision Matrix Estimator]
\label{thm:unbiased-precision}
Let $\mathcal{B}$ be a mini-batch of size $B$ sampled uniformly without replacement
from $\{1,\dots,N\}$. Define
\begin{align*}
\hat{\bs\Lambda}_{\bs\lambda,\bs\phi}^{(\mathcal{B})}
= \frac{N}{B} \sum_{i \in \mathcal{B}} \X_i^\top \L_i \L_i^\top \X_i + \K_{\bs\lambda}.
\end{align*}
Then
\begin{align*}
\mathbb{E}_{\mathcal{B}}
\!\left[ \hat{\bs\Lambda}_{\bs\lambda,\bs\phi}^{(\mathcal{B})} \right]
= \bs\Lambda_{\bs\lambda,\bs\phi}.
\end{align*}
\end{theorem}

\begin{proof}
\begin{align*}
\mathbb{E}_{\mathcal{B}} \left[ \hat{\bs\Lambda}_{\bs\lambda,\bs\phi}^{(\mathcal{B})} \right]
&= \mathbb{E}_{\mathcal{B}} \left[ \frac{N}{B}
\sum_{i \in \mathcal{B}}
\X_i^\top \L_i \L_i^\top \X_i \right] + \K_{\bs\lambda}  \\
&= \frac{N}{B} \cdot \frac{B}{N}
\sum_{i=1}^N \X_i^\top \L_i \L_i^\top \X_i + \K_{\bs\lambda} \\
&= \sum_{i=1}^N \X_i^\top \L_i \L_i^\top \X_i + \K_{\bs\lambda} \\
&= \bs\Lambda_{\bs\lambda,\bs\phi}.
\end{align*}
\end{proof}

\begin{theorem}[Unbiasedness of the Natural Mean Estimator]
\label{thm:unbiased-natural-mean}
Let
\begin{align*}
\hat{\bs\Lambda}_{\bs\lambda,\bs\phi}^{(\mathcal{B})}\,
\hat{\tilde{\bs\beta}}_{\bs\lambda,\bs\phi}^{(\mathcal{B})}
=
\frac{N}{B} \sum_{i \in \mathcal{B}}
\X_i^\top \L_i \L_i^\top \bs\mu_i .
\end{align*}
Then
\begin{align*}
\mathbb{E}_{\mathcal{B}}
\!\left[
\hat{\bs\Lambda}_{\bs\lambda,\bs\phi}^{(\mathcal{B})}\,
\hat{\tilde{\bs\beta}}_{\bs\lambda,\bs\phi}^{(\mathcal{B})}
\right]
=
\bs\Lambda_{\bs\lambda,\bs\phi}\,
\tilde{\bs\beta}_{\bs\lambda,\bs\phi}.
\end{align*}
\end{theorem}

\begin{proof}
\begin{align*}
\mathbb{E}_{\mathcal{B}}
\left[
\hat{\bs\Lambda}_{\bs\lambda,\bs\phi}^{(\mathcal{B})}\,
\hat{\tilde{\bs\beta}}_{\bs\lambda,\bs\phi}^{(\mathcal{B})}
\right]
&= \frac{N}{B} \cdot
\mathbb{E}_{\mathcal{B}} \left[
\sum_{i \in \mathcal{B}}
\X_i^\top \L_i \L_i^\top \bs\mu_i
\right] \\
&= \frac{N}{B} \cdot \frac{B}{N}
\sum_{i=1}^N \X_i^\top \L_i \L_i^\top \bs\mu_i \\
&= \bs\Lambda_{\bs\lambda,\bs\phi}\,
\tilde{\bs\beta}_{\bs\lambda,\bs\phi}.
\end{align*}
\end{proof}

We next analyze the effect of mini-batching on the ELBO.

\paragraph{Full-data ELBO.}
The ELBO for the complete dataset is
\begin{align}
\text{ELBO}(\bs\lambda,\bs\phi)
=
\mathbb{E}_{q(\bs\beta \mid \bs\lambda,\X,\bs\phi)}
\!\left[
\log p(\bs\beta)
+ \sum_{i=1}^N \log p(y_i \mid \bs\beta)
- \log q(\bs\beta \mid \bs\lambda,\X,\bs\phi)
\right].
\label{eq:full-elbo-app}
\end{align}

\paragraph{Stochastic mini-batch ELBO.}
When constructing the variational distribution using only the mini-batch
$\X^{\mathcal{B}}$, the stochastic ELBO becomes
\begin{align}
\widehat{\text{ELBO}}^{(\mathcal{B})}
&=
\mathbb{E}_{q(\bs\beta \mid \bs\lambda,\X^{\mathcal{B}},\bs\phi)}
\!\left[
\log p(\bs\beta)
+ \frac{N}{B} \sum_{i \in \mathcal{B}} \log p(y_i \mid \bs\beta)
- \log q(\bs\beta \mid \bs\lambda,\X^{\mathcal{B}},\bs\phi)
\right].
\label{eq:stochastic-elbo-app}
\end{align}

\begin{theorem}[Bias of the Mini-batch ELBO Estimator]
\label{thm:elbo-bias}
Suppose that the variational distribution $q(\bs\beta \mid \bs\lambda,\X^{\mathcal{B}},\bs\phi)$
depends on the sampled mini-batch $\mathcal{B}$. Then, in general,
\begin{align*}
\mathbb{E}_{\mathcal{B}}
\!\left[
\widehat{\text{ELBO}}^{(\mathcal{B})}
\right]
\neq
\text{ELBO}(\bs\lambda,\bs\phi),
\end{align*}
where $\text{ELBO}(\bs\lambda,\bs\phi)$ is defined in~\eqref{eq:full-elbo-app}
and $\widehat{\text{ELBO}}^{(\mathcal{B})}$ in~\eqref{eq:stochastic-elbo-app}.
\end{theorem}

\begin{proof}
Let
$$
q_{\mathcal B}(\bs\beta) = q(\bs\beta\mid\bs\lambda,\X_{\mathcal B},\bs\phi)
$$
denote the variational distribution constructed from the random mini-batch $\mathcal B$. The stochastic mini-batch ELBO is
$$
\widehat{\operatorname{ELBO}}_{\mathcal B} = E_{q_{\mathcal B}}\left[\frac{N}{B}\sum_{i\in\mathcal B} \log p(y_i\mid\bs\beta) + \log p(\bs\beta\mid\bs\lambda) - \log q_{\mathcal B}(\bs\beta) \right].
$$
For a fixed value of $\bs\beta$, uniform sampling of $\mathcal B$ and the factor $N/B$ yield
$$
E_{\mathcal B}\left[\frac{N}{B}\sum_{i\in\mathcal B} \log p(y_i\mid\bs\beta)\right] = \sum_{i=1}^N\log p(y_i\mid\bs\beta).
$$
However, this pointwise unbiasedness does not imply unbiasedness of the mini-batch ELBO, because the distribution with respect to which the expectation is taken also depends on $\mathcal B$. In general,
$$
E_{\mathcal B} \left[E_{q_{\mathcal B}}\left[\frac{N}{B}\sum_{i\in\mathcal B}\log p(y_i\mid\bs\beta)\right]\right] \neq E_{q} \left[\sum_{i=1}^N\log p(y_i\mid\bs\beta)\right],
$$
where
$$
q(\bs\beta) = q(\bs\beta\mid\bs\lambda,\X,\bs\phi)
$$
is the full-data variational distribution. For the same reason, in general,
$$
E_{\mathcal B}\left[E_{q_{\mathcal B}}[\log p(\bs\beta\mid\bs\lambda)]\right] \neq E_q[\log p(\bs\beta\mid\bs\lambda)]
$$
and
$$
E_{\mathcal B}\left[E_{q_{\mathcal B}}[\log q_{\mathcal B}(\bs\beta)]\right] \neq E_q[\log q(\bs\beta)].
$$
Thus, although the sufficient statistics used to construct $q_{\mathcal B}$ are unbiased estimators of their full-data counterparts, the nonlinear dependence of the variational distribution on the sampled mini-batch means that the resulting stochastic ELBO is, in general, biased:
$$
E_{\mathcal B}\left[\widehat{\operatorname{ELBO}}_{\mathcal B}\right] \neq \operatorname{ELBO}(\bs\lambda,\bs\phi).
$$
\end{proof}

\clearpage

\subsection{Asymptotic complexity}

\autoref{tab:asymptotic-complexity} compares the per-iteration asymptotic complexity of our three SVI variants for estimating the regression coefficients $\bs\beta$ with INLA and the No-U-Turn Sampler (NUTS). For compactness, we refer to the three SVI variants as \texttt{svi\_amortized}, \texttt{svi\_amortized\_bd} (block-diagonal), and \texttt{svi\_dense}.

\begin{table}[tbhp]
\centering
\begin{tabular}{@{}cc@{}}
\toprule
\textbf{Model}      & \textbf{Asymptotic complexity}                                                                                                                                                                                                                                                                                                                                                                                                                                                                      \\ \midrule
\texttt{svi\_amortized}     & $\mathcal{O}(\underbrace{\vphantom{\sum_{p = 1}^P}Q^3 + NG}_\text{Construction of $q$} + \underbrace{\vphantom{\sum_{p = 1}^P}SNQ}_{\ln p(\y | \bs\beta, \X)} + \underbrace{S\sum_{p = 1}^P \sum_{j = 1}^{J_p} Q^2_{p, j}}_{\ln p(\bs\beta | \bs\lambda)} + \underbrace{\vphantom{\sum_{p = 1}^P}SQ^2}_{\ln q(\bs\beta | \bs\lambda, \X, \bs\phi)})$                                                                                                    \\
\texttt{svi\_amortized\_bd} & $\mathcal{O}(\underbrace{Q^2 + \sum_{p = 1}^P \sum_{j = 1}^{J_p} Q^3_{p, j} + N\sum_{p = 1}^P \sum_{j = 1}^{J_p} G_{p, j}}_\text{Construction of $q$} + \underbrace{SNQ}_{\ln p(\y | \bs\beta, \X)} + \underbrace{S\sum_{p = 1}^P \sum_{j = 1}^{J_p} Q^2_{p, j}}_{\ln p(\bs\beta | \bs\lambda)} + \underbrace{\vphantom{\sum_{p = 1}^P}SQ^2}_{\ln q(\bs\beta | \bs\lambda, \X, \bs\phi)})$                                                                 \\
\texttt{svi\_dense}  & $\mathcal{O}(\underbrace{\vphantom{\sum_{p = 1}^P}Q^2}_\text{Construction of $q$} + \underbrace{\vphantom{\sum_{p = 1}^P}SNQ}_{\ln p(\y | \bs\beta, \X)} + \underbrace{S\sum_{p = 1}^P \sum_{j = 1}^{J_p} Q^2_{p, j}}_{\ln p(\bs\beta | \bs\lambda)} + \underbrace{\vphantom{\sum_{p = 1}^P}SQ^2}_{\ln q(\bs\beta | \bs\lambda, \X, \bs\phi)})$                                                                                                            \\
INLA           & $\mathcal{O}(\underbrace{\vphantom{\sum_{p = 1}^P}Q^3}_\text{Factorization of the precision matrix} + \underbrace{\vphantom{\sum_{p = 1}^P}NQ}_{\ln p(\y | \bs\beta, \X)} + \underbrace{\sum_{p = 1}^P \sum_{j = 1}^{J_p} Q^2_{p, j}}_{\ln p(\bs\beta | \bs\lambda)})$                                                                                                                                                                                     \\
NUTS           & $\mathcal{O}(\underbrace{\sum_{p = 1}^P 2^{D_{p}}}_\text{Leapfrog steps} \underbrace{\vphantom{\sum_{p = 1}^P}Q_{p}}_\text{Gradient of $\ln p(\y, \bs\beta_{p} | \bs\lambda_{p}, \X_{p})$} + \underbrace{\vphantom{\sum_{p = 1}^P}NQ}_{\ln p(\y | \bs\beta, \X)} + \underbrace{\sum_{p = 1}^P Q^2_{p}}_{\ln p(\bs\beta | \bs\lambda)})$ \\ \bottomrule
\end{tabular}
\caption{Asymptotic complexity of different SVI models compared with INLA and the No-U-Turn Sampler (NUTS). For the complexity analysis, we consider the full-batch setting $B=N$. The symbols $G$ and $G_{p, j}$ represent the number of atomic operations for a forward pass of the neural network. For the amortized variants, we report the dominant cost of constructing the variational distribution. The cost of accumulating the Gaussian sufficient statistics is omitted, since under the neural architectures considered here it is dominated by the cost of the neural-network forward evaluations. Consider the case where the neural network used in \texttt{svi\_amortized} has a single hidden layer with $M$ neurons. The total number of operations required to compute both the mean and precision matrix is $(Q + J + 1) \cdot M + M + M \cdot \frac{P(P + 1)}{2} + \frac{P(P + 1)}{2}$, where $J=\sum_{p=1}^P J_p$ is the total number of smoothing parameters. Assuming $M \approx Q$, this remains computationally efficient. In contrast, for \texttt{svi\_amortized\_bd}, the total number of operations is $\sum_{p=1}^P \sum_{j=1}^{J_p} \left[ (Q_{p,j} + 2) \cdot M_{p,j} + M_{p,j} + 2 M_{p,j} + 2 \right]$, where $M_{p,j}$ is the number of neurons in the hidden layer of the neural network modeling $\tilde{\bs\beta}_{p,j}$ and the block precision matrix $\bs\Lambda_{p,j}(\lambda_{p,j}, \bs\phi_{p,j})$. Due to the QR-reparametrization applied to each design matrix $\X_i$, the precision matrix in INLA becomes dense, leading to a cubic complexity for its factorization. We employ one No-U-turn sampler (NUTS) for each block $\beta_{p}$ of the regression coefficients.}
\label{tab:asymptotic-complexity}
\end{table}

\clearpage

\subsection{Monitoring Convergence via a Moving--Average ELBO}\label{subsec:ma_elbo}

Stochastic variational inference produces inherently noisy estimates of the Evidence Lower Bound due to both mini–batching and Monte Carlo sampling. Moreover, computing the full ELBO at every iteration would be prohibitively expensive for all models considered in this work. As shown in Table~\ref{tab:asymptotic-complexity}, evaluating the ELBO involves terms with quadratic and cubic dependence on $Q$, and in some cases additional dependence on $N$, $P$, and $G$, making repeated full ELBO computations infeasible in practice. Consequently, the full ELBO is too costly to compute at each iteration, whereas the mini–batch ELBO is too noisy to serve as a reliable convergence diagnostic.

To obtain a stable and computationally efficient measure of convergence, we track an exponential moving average of the noisy ELBO values. We refer to this diagnostic as the \emph{moving--average ELBO} (MA--ELBO). Let $\widehat{\mathrm{ELBO}}_{t}$ denote the stochastic ELBO estimate at iteration $t$, computed from the current mini–batch and a single Monte Carlo sample from the variational distribution. The MA--ELBO is defined recursively as
\begin{align}
\mathrm{MA\text{-}ELBO}_{t}
&= (1 - \rho)\,\mathrm{MA\text{-}ELBO}_{t-1}
  + \rho\,\widehat{\mathrm{ELBO}}_{t},
\label{eq:ma_elbo}
\end{align}
where $\rho \in (0,1)$ is a smoothing parameter. Smaller values of $\rho$ yield stronger smoothing but slower reaction to changes in the underlying ELBO, whereas larger values track recent iterations more closely. In all numerical experiments we select $\rho = 0.05$, which provides a good balance between stability and responsiveness.

The MA--ELBO does not alter the optimization objective; it is used solely for monitoring convergence. In practice, the recursion in~\eqref{eq:ma_elbo} produces a low–variance, interpretable curve whose evolution is substantially easier to assess than that of the raw mini–batch ELBO. This is particularly relevant for amortized variational distributions $q(\boldsymbol{\beta}\mid\boldsymbol{\lambda},\mathbf{X},\boldsymbol{\phi})$, whose local structure depends on the current mini–batch and therefore exhibits high stochastic fluctuations. Similar moving–average smoothing of noisy ELBO traces has been used in other SVI applications; for example, \citep{li2021bayesian} employ a moving–average ELBO with a finite window to diagnose convergence in a Bayesian LSTM model.

From a theoretical perspective, moving–average gradient estimators are known to reduce variance in non–convex stochastic optimization and to provide sufficiently accurate gradient information for convergence of adaptive optimizers to stationary points \citep{Guo2021OnSM}. In our context, the MA--ELBO plays an analogous role: it yields a stable, monotone–in–expectation diagnostic that allows us to determine whether the optimization of the variational parameters has effectively converged.

Throughout both the simulation study and the application, we therefore report the MA--ELBO rather than the raw ELBO. Convergence is assessed by visual inspection of the MA--ELBO trace.

\clearpage

\subsubsection{Simulation model architecture and implementation details}
\label{app:architecture-sim}

This subsection documents the implementation choices used in the simulation study of Section~\ref{section:Simulation}, including priors, variational families, neural networks, sampling settings, optimizers, spline specifications, and the MCMC benchmark.

\paragraph{Priors.}
All smoothing parameters are represented on the log scale,
$$
\tau_{p,j}=\log\lambda_{p,j},
$$
where $\lambda_{p,j}^2$ denotes the corresponding precision, and endowed with independent Exp-Gamma priors with concentration parameter $1.0$ and rate $0.01$.  
The probability density function of the Exp-Gamma distribution with concentration parameter $\alpha > 0$ and rate parameter $\beta > 0$ is
\begin{align}
    p(x)
    = \frac{e^{x\alpha} e^{-\beta e^{x}}}{\Gamma(\alpha)\,\beta^{-\alpha}},
    \label{eq:Exp-Gamma}
\end{align}
which induces a prior on the smoothing variances that favours moderate smoothness while allowing for flexible deviations when strongly supported by the data.

\paragraph{Variational distributions and Monte Carlo sampling.}
Each smoothing parameter $\tau_{p,j}$ is assigned a Gaussian variational distribution.  
We use the following numbers of Monte Carlo samples: $S_{\beta} = 32$ and $S_\tau = 1$.

\paragraph{Neural network architecture.}
All neural networks used in the amortized variational distributions consist of a single hidden layer with $16$ neurons and ReLU activation.

\paragraph{Optimizers and training regime.}
Each SVI model is trained for $8000$ epochs using the Adam optimizer~\cite{kingma2014adam} with learning rate 0.05. All simulation experiments are performed in the full-batch setting, i.e., $B=N$, such that every optimization step uses the complete data set.

\paragraph{Spline specifications.}
For the Bernoulli simulations, the nonlinear effects of $x_{1}$ and $x_{2}$ are modeled using cubic P-splines with 10 knots and a second-order random walk penalty.  

For the Gamma simulations, the covariate $x$ influences both the mean and the variance through cubic P-splines with 10 knots and a second-order random walk prior.

\paragraph{MCMC benchmark.}
The gold-standard posterior is obtained using NUTS sampling implemented in the Liesel framework~\cite{liesel}. We apply NUTS with 8000 warmup iterations, followed by 4000 posterior and 4000 terminal samples.

\paragraph{Computational environment.}
All simulations were conducted on a 2021 MacBook Pro with an Apple M1 Pro CPU, 32GB RAM, macOS Tahoe 26.1, and a 512GB SSD.

\medskip

These settings ensure that all SVI variants are trained under consistent conditions and provide a high-accuracy MCMC reference for Wasserstein distance comparisons and coverage analyses in Section~\ref{section:Simulation}.

\clearpage

\subsection{Additional figures and tables}

\begin{figure}[ht]
    \centering
    \includegraphics[width=\textwidth]{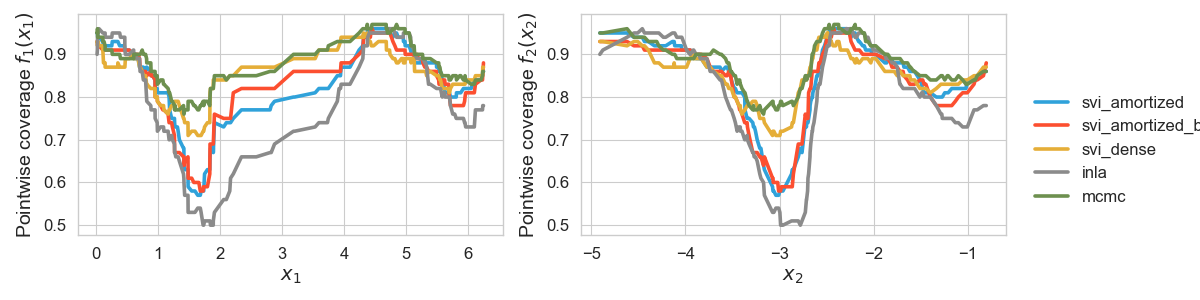} 
    \caption{Logistic model - Coverage probabilities of the nominal 95\% credible intervals across the domain of the two smooth effects. The left panel corresponds to the effect of $x_{1}$, while 
    the right panel shows the effect of $x_{2}$.}
    \label{fig:bernoulli-coverage-probabilities}
\end{figure}

\begin{figure}[ht]
    \centering
    \includegraphics[width=\textwidth]{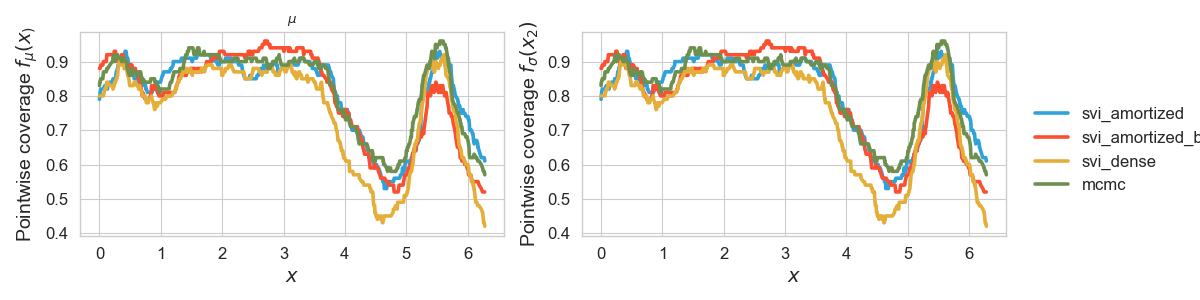} 
    \caption{Gamma model - Coverage probabilities for the two smooth effects. 
    The left panel corresponds to the mean parameter $\mu$, while the right panel 
    corresponds to the variance parameter $\sigma^{2}$.}
    \label{fig:gamma-coverage-probabilities}
\end{figure}

\subsection{Applications}

\subsection{Neural architectures and optimization details}\label{app:architecture}

This section describes the neural architectures, hyperpriors, sampling settings, and optimization scheme used in the amortized and block–diagonal stochastic variational inference models.

\paragraph{Hyperpriors.}
Each smooth component is associated with a log–smoothing parameter $\tau$, endowed with an independent $\mathrm{ExpGamma}(1.0,\, 0.01)$ hyperprior.  
This prior encourages moderate smoothness while allowing flexible departures when supported by the data.

\paragraph{Optimization.}
All variational parameters are optimized using the Adam optimizer with learning rate $0.001$, combined with global–norm gradient clipping (threshold $1.0$) and automatic removal of NaNs during optimization. We use mini-batches of size $B=4096$ at each optimization step.

\paragraph{Monte Carlo sample sizes.}
We use a single Monte Carlo sample for the smoothing parameters ($S_{\tau}=1$) and $64$ Monte Carlo samples for the regression coefficients ($S_{\beta}=64$) at each iteration.

\paragraph{Amortized SVI network.}
The fully amortized variational distribution uses a single neural network shared across all smooth effects.  
This network consists of one hidden layer with $128$ neurons and ReLU activation, and outputs the local Gaussian parameters required for the amortized SVI updates.

\paragraph{Block–diagonal SVI networks.}
For the block–diagonal variational distribution, the predictor is decomposed into independent blocks corresponding to model effects, and a separate neural network is used for each block.  
Every such network has a single hidden layer with $16$ neurons and ReLU activation, and outputs the mean and log-Cholesky parameters of the corresponding Gaussian block in the block–diagonal variational approximation.

\begin{figure}[t]
  \centering
  \includegraphics[width=0.5\textwidth]{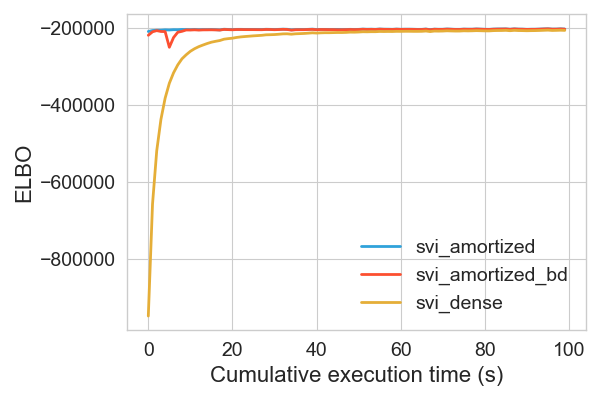}
  \caption{Estimated evidence lower bound (ELBO) for the three SVI methods as a function of optimization time. The proposed amortized approaches (\texttt{svi\_amortized} in blue and \texttt{svi\_amortized\_bd} in red) reach higher ELBO values than the dense baseline (\texttt{svi\_dense}, yellow) while requiring less computation. The block-diagonal version shows a short transient phase with slightly more variability at the beginning, but stabilizes at essentially the same ELBO level as the fully amortized model.}
  \label{fig:elbo_nyc}
\end{figure}

\subsection{A second case study}\label{app:second_application}

We now assess the proposed variational approximations on the patent citations data set from \citet{jerak2006modeling}.  The data contain patents from biotechnology/pharmaceutics and semiconductor/computer technologies; an overview of all variables is reported in \autoref{tab:patents-citationts}. Following \citet{jerak2006modeling}, we remove about 1\% of observations with extreme values in some covariates, which leaves 4{,}804 patents. Our aim is to predict the number of forward citations using the explanatory variables, with special emphasis on the continuous covariate ``number of claims''.

We specify a Negative Binomial model for the count response and model both the mean $\mu>0$ and the dispersion $\delta>0$ with additive predictors. For both parameters we use the softplus link
\begin{align*}
g(x) = \log\bigl(1 + \exp(x)\bigr),
\end{align*}
so that $\mu_i = g(\eta_{\mu,i})$ and $\delta_i = g(\eta_{\delta,i})$. In particular, we let both predictors depend on all available covariates, that is
\begin{align*}
  \eta_{\ell,i}
    &= \beta_{\ell,0}
       + f_{\ell,\text{year}}(\text{year}_i)
       + f_{\ell,\text{ncountry}}(\text{ncountry}_i)
       + f_{\ell,\text{nclaims}}(\text{nclaims}_i) \\
    &\quad
       + \text{opp}_i + \text{biopharm}_i + \text{ustwin}_i
       + \text{patus}_i + \text{patgsgr}_i,
\end{align*}
for $\ell \in \{\mu,\delta\}$.  
All smooth functions $f_{\bullet,\bullet}$ are represented by cubic P-splines with ten knots, while the remaining covariates enter the model linearly.

As discussed in \citet{Klein2015Bayesian}, this data set is challenging from an inferential perspective. The number of citations is highly overdispersed, but there is little evidence of zero inflation. The corresponding likelihood is rather flat, which leads to marginal posterior distributions for the regression coefficients that deviate from normality and induces non-stationary MCMC paths when conventional Inverse-Gamma priors are used for the variance components. \citet{Klein2016Scale} alleviate these problems by adopting scale-dependent priors, which stabilise the MCMC chains and reduce the amount of manual tuning required.

In our analysis we follow the prior specification of \citet{Klein2016Scale} and compare four inference schemes: the two Gaussian variational approximations introduced in \autoref{subsec:Amortized Variational Distribution} and \autoref{subsec:Amortized Variational Distribution - Block Diagonal}
(\texttt{svi\_amortized}, \texttt{svi\_amortized\_bd}, the dense SVI baseline introduced in Section \texttt{svi\_dense}) introduced in \autoref{section:Simulation} and an MCMC baseline. The MCMC sampler is implemented in Liesel~\citep{liesel} using one NUTS kernel for each block of regression coefficients and a separate kernel for all smoothing parameters.

\paragraph{Neural architectures and optimization details.}

For the three SVI variants we use the same hyperpriors as in \citet{Klein2016Scale}; a detailed description is provided in \autoref{app:architecture}. All variational parameters are optimized with the Adam optimizer with learning rate $0.01$, global–norm gradient clipping (threshold $1.0$), and automatic removal of numerical instabilities (NaNs) during optimization. We draw a single Monte Carlo sample for the smoothing parameters ($S_{\tau}=1$) and $64$ Monte Carlo samples for the regression coefficients ($S_{\beta}=64$) at each iteration. Gradients are always computed on the full data set, that is, we do not use any form of mini-batching.

The fully amortized variational approximation employs a single neural network shared across all smooth effects. This network has one hidden layer with $128$ neurons and ReLU activation, and outputs the local Gaussian parameters required for the amortized SVI updates. For the block–diagonal variational approximation, the predictor is decomposed into independent blocks corresponding to model effects, and a separate neural network is used for each block. Every such network has a single hidden layer with $16$ neurons and ReLU activation, and outputs the mean and log-Cholesky parameters of the corresponding Gaussian block.

\begin{figure}[tbhp]
    \centering
    \includegraphics[width=\textwidth]{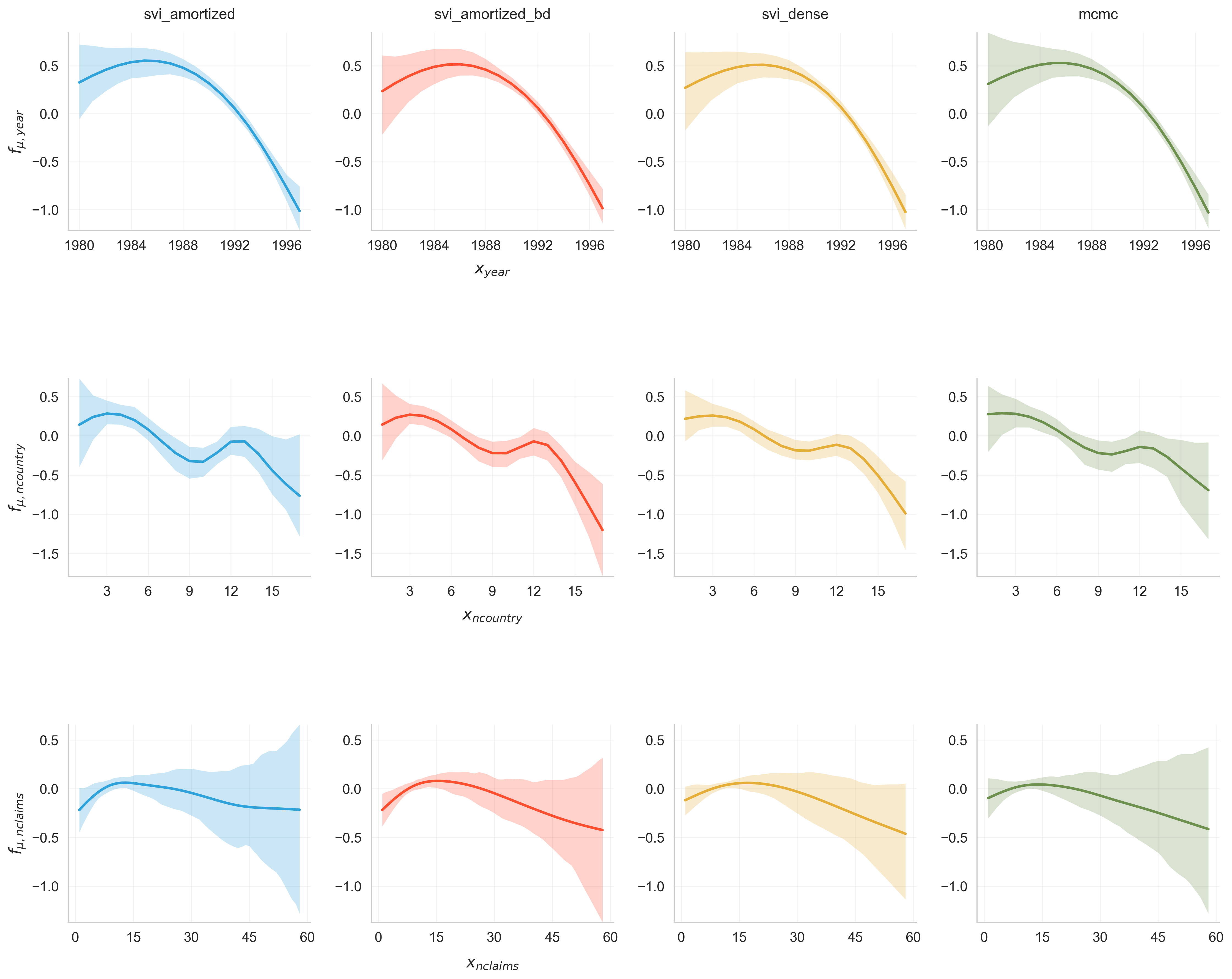}
    \caption{Patents citations data set – Posterior mean and credible
      intervals of the non-linear effects of \textit{year},
      \textit{ncountry}, and \textit{nclaims} on the predictor for the
      mean for the four inference schemes: \texttt{svi\_amortized},
      \texttt{svi\_amortized\_bd}, \texttt{svi\_dense}, and MCMC.}
    \label{fig:patents-citationts-effects-mu}
\end{figure}

\autoref{fig:patents-citationts-effects-mu} shows the posterior mean and credible intervals of the non-linear effects of \textit{year}, \textit{ncountry}, and \textit{nclaims} on the predictor for the mean for each method. All approaches yield very similar shapes. For the year effect, the estimated curves agree on a nearly constant effect in the early 1980s followed by a pronounced decline in expected citations for more recent patents. The effect of the number of countries exhibits a weakly non-linear pattern, with moderate changes around the mode that are captured in a consistent way by all four methods. The effect of the number of claims is predominantly negative, indicating that patents with many claims tend to receive fewer citations, again with close agreement between variational approximations and MCMC. The variational methods typically produce slightly narrower credible bands than MCMC, which is expected from an approximation that shrinks the tails of the posterior.

\begin{figure}[tbhp]
    \centering
    \includegraphics[width=\textwidth]{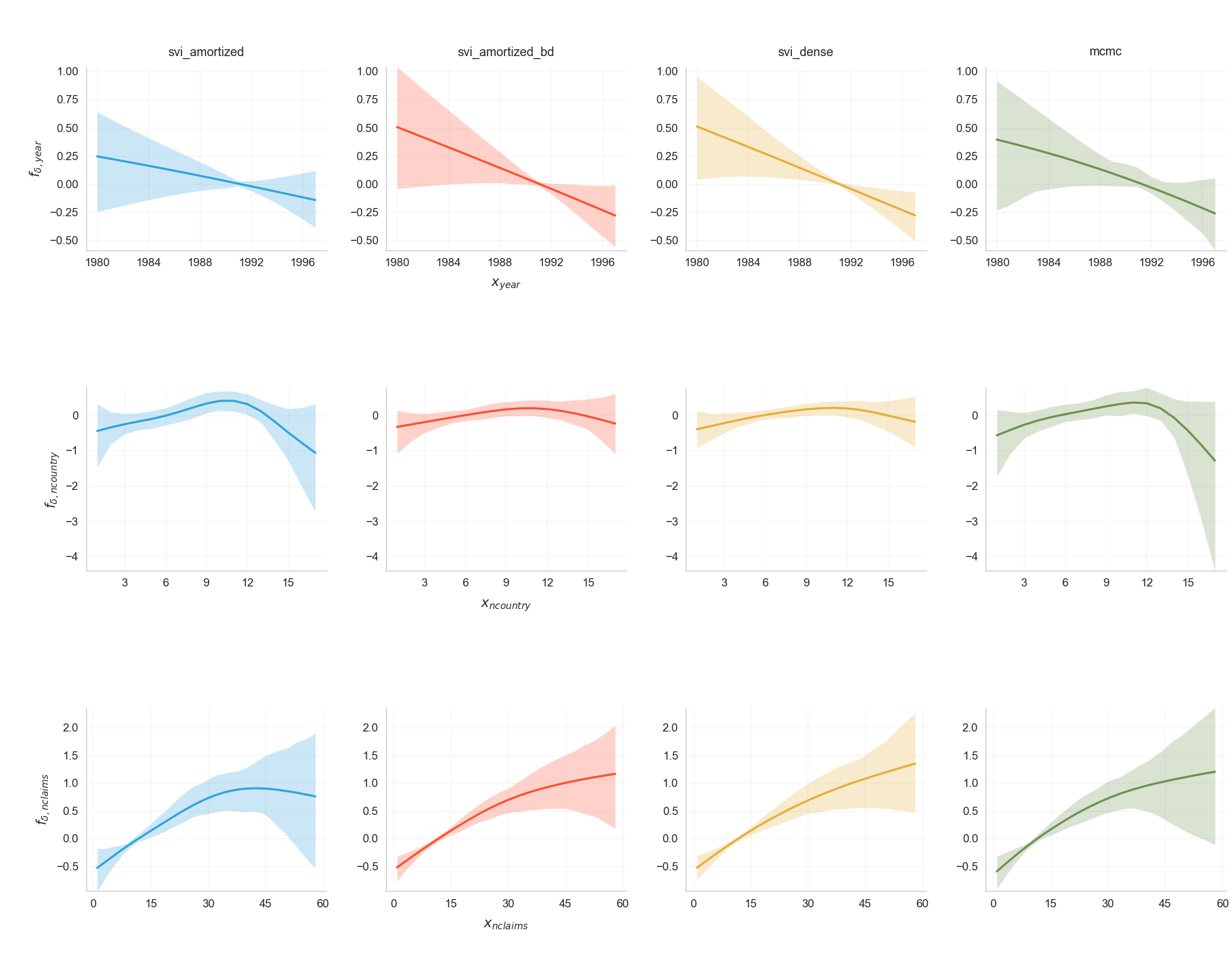}
    \caption{Patents citations data set – Posterior mean and credible
      intervals of the non-linear effects of \textit{year},
      \textit{ncountry}, and \textit{nclaims} on the dispersion
      predictor for the four inference schemes: \texttt{svi\_amortized},
      \texttt{svi\_amortized\_bd}, \texttt{svi\_dense}, and MCMC.}
    \label{fig:patents-citationts-effects-delta}
\end{figure}

\autoref{fig:patents-citationts-effects-delta} displays the corresponding estimated effects on the dispersion predictor.  
All methods agree that the dispersion decreases over time, implying that citation counts for more recent patents are less variable than those for older ones. The effect of the number of countries is again mildly non-linear and very similar across methods, with higher variability for patents filed in many countries. Finally, the dispersion clearly increases with the number of claims, which means that patents with many claims exhibit a wider spread in their citation counts. As for the mean structure, the three variational approximations closely track the MCMC estimates, and the main differences are confined to regions where data are sparse and the credible intervals are wide for all methods.

In summary, although the patent citations data set is known to cause difficulties for MCMC due to the flat likelihood and the resulting non-standard posterior shapes, the Gaussian variational distributions considered here recover non-linear effects that are very close to those obtained by MCMC, both for the mean and for the dispersion.  
This empirical agreement supports the use of the proposed variational approximations in complex additive count models, where they provide computationally efficient inference without a substantial loss in accuracy.

\begin{table}[ht]
\centering
\begin{tabular}{@{}ccccc@{}}
\toprule
\textbf{Variable} & 
\textbf{Description} & 
\begin{tabular}[c]{@{}c@{}}
  \textbf{Mean /}\\
  \textbf{Frequency in \%}
\end{tabular} &
\textbf{Std-dev.} & 
\textbf{Min/max} \\ \midrule
\textit{opp}      & \begin{tabular}[c]{@{}c@{}}Patent opposition\\ 1 = yes\\ 0 = no\end{tabular}                                            & \begin{tabular}[c]{@{}c@{}}41.09\\ 58.91\end{tabular} &                   &                  \\
\textit{biopharm} & \begin{tabular}[c]{@{}c@{}}Patent from biotech/pharm sector\\ 1 = yes\\ 0 = no\end{tabular}                             & \begin{tabular}[c]{@{}c@{}}43.78\\ 56.22\end{tabular} &                   &                  \\
\textit{ustwin}   & \begin{tabular}[c]{@{}c@{}}US twin patent exists\\ 1 = yes\\ 0 = no\end{tabular}                                        & \begin{tabular}[c]{@{}c@{}}61.28\\ 38.72\end{tabular} &                   &                  \\
\textit{patus}    & \begin{tabular}[c]{@{}c@{}}Patent holder from the USA\\ 1 = yes\\ 0 = no\end{tabular}                                   & \begin{tabular}[c]{@{}c@{}}31.18\\ 68.82\end{tabular} &                   &                  \\
\textit{patgsgr}  & \begin{tabular}[c]{@{}c@{}}Patent holder from Germany, \\ Switzerland, or Great Britain\\ 1 = yes\\ 0 = no\end{tabular} & \begin{tabular}[c]{@{}c@{}}23.77\\ 76.23\end{tabular} &                   &                  \\
\textit{year}     & Grant year                                                                                                              &                                                       &                   & 1980 - 1997      \\
\textit{ncit}     & \begin{tabular}[c]{@{}c@{}}Number of citations for the\\ patent\end{tabular}                                            & 1.52                                                  & 2.25              & 0 - 14           \\
\textit{ncountry} & \begin{tabular}[c]{@{}c@{}}Number of designated states\\ for the patent\end{tabular}                                    & 2.79                                                  & 4.12              & 1 - 17           \\
\textit{nclaims}  & Number of claims                                                                                                        & 12.48                                                 & 8.57              & 1 - 58           \\ \bottomrule
\end{tabular}
\caption{Patent citations: description of variables including summary statistics}
\label{tab:patents-citationts}
\end{table}

\end{document}